\documentclass[11pt,oneside]{article}

\usepackage{palatino}
\usepackage[margin=1in]{geometry}
\usepackage{color,xcolor,soul}
\usepackage[abspage,user,savepos]{zref}
\usepackage[colorlinks, citecolor=blue,linkcolor=blue]{hyperref}  
\usepackage{natbib}
\usepackage{fp}
\usepackage{framed}
\usepackage{cuted,balance}
\usepackage{setspace}
\usepackage{subcaption}
\usepackage{lipsum}
\usepackage{amsmath}
\usepackage{amssymb,bm}
\usepackage{mathtools}
\usepackage{amsopn} 
\usepackage{amsfonts}
\usepackage{dashrule}
\usepackage{multirow}
\usepackage{microtype}
\usepackage{multicol}  
\usepackage{color}      
\usepackage{subcaption}   
\usepackage{booktabs}   
\usepackage{epstopdf}
\usepackage{algorithm} 
\usepackage{algorithmic}
\usepackage{dsfont}
\usepackage{makecell}
\usepackage{graphicx}
\usepackage{cleveref}
\usepackage{tikz}
\usepackage{graphicx} 
\usepackage{epsfig} 
\usepackage{mathtools}
\usepackage{dsfont,amsmath} 
\usepackage{amsthm}
\usepackage{amssymb}  
\usepackage{algorithm,algorithmic}

\allowdisplaybreaks
\newtheorem{theorem}{Theorem}
\newtheorem{lemma}{Lemma}
\newtheorem{corollary}{Corollary}
\newtheorem{claim}{Claim}
\crefname{claim}{Claim}{Claims}
\newtheorem{definition}{Definition}
\newtheorem{remark}{Remark}
\newtheorem{proposition}{Proposition}
\newtheorem{assumption}{Assumption}
\crefname{assumption}{Assumption}{Assumption}

\DeclareMathOperator*{\argmax}{argmax}

\newcommand{\wbar}[1]{\overline{#1}}
\newcommand{\what}[1]{\widehat{#1}}

\newcommand{\TV}[1]{\left\|#1\right\|_{\mathrm{TV}}}

\newcommand{\be}{\begin{equation}}
\newcommand{\ee}{\end{equation}}
\newcommand{\E}{\mathrm{E}}

\newcommand{\nn}{\nonumber}
\newcommand{\bs}[1]{#1}
\newcommand{\br}{\underline{\mathrm{br}}}

\newcommand{\ua}{\underline{a}}
\newcommand{\uP}{\underline{P}}

\newcommand{\piave}{\pi^\mathrm{ave}}
\newcommand{\pifree}{\pi^\mathrm{free}}
\newcommand{\ane}{\mathrm{a.n.e.}}

\makeatletter
\newcommand{\myitem}[1]{%
\item[#1]\protected@edef\@currentlabel{#1}%
}
\makeatother

\newcommand{\oalpha}{\overline{\alpha}}

\newcommand{\oA}{\overline{A}}
\newcommand{\uA}{\underline{A}}
\newcommand{\onu}{\overline{\nu}}
\newcommand{\unu}{\underline{\nu}}

\newcommand{\oK}{\overline{K}}
\newcommand{\uK}{\underline{K}}

\begin{document}

\date{}

\title{\LARGE \bf
Logit-Q Dynamics for Efficient Learning \\in Stochastic Teams
}

\author{Ahmed Said Donmez\thanks{A. S. Donmez and O. Unlu made equal contributions.}\textsuperscript{\hspace{1.3mm}}\thanks{Department of Electrical and Electronics Engineering, Bilkent University, Ankara, T\"{u}rkiye 
  (said.donmez@bilkent.edu.tr, sayin@ee.bilkent.edu.tr).} \and
  Onur Unlu\footnotemark[1]\textsuperscript{\hspace{1.3mm}}\thanks{Department of Electrical and Computer Engineering, Cornell University, Ithaca, NY (ou25@cornell.edu).} \and  Muhammed O. Sayin\footnotemark[2]}

\maketitle

\begin{abstract}
    We present a new family of logit-Q dynamics for efficient learning in stochastic games by combining the log-linear learning (also known as logit dynamics) for the repeated play of normal-form games with Q-learning for unknown Markov decision processes within the auxiliary stage-game framework. In this framework, we view stochastic games as agents repeatedly playing some stage game associated with the current state of the underlying game while the agents' Q-functions determine the payoffs of these stage games. We show that the logit-Q dynamics presented reach (near) efficient equilibrium in stochastic teams with unknown dynamics and quantify the approximation error. We also show the rationality of the logit-Q dynamics against agents following pure stationary strategies and the convergence of the dynamics in stochastic games where the stage-payoffs induce potential games, yet only a single agent controls the state transitions beyond stochastic teams. The key idea is to approximate the dynamics with a fictional scenario where the Q-function estimates are stationary over epochs whose lengths grow at a sufficiently slow rate. We then couple the dynamics in the main and fictional scenarios to show that these two scenarios become more and more similar across epochs due to the vanishing step size and growing epoch lengths.
    \end{abstract}
    
    \section{Introduction}
    
    Stochastic games, introduced by \cite{ref:Shapley53}, generalize Markov decision processes to non-cooperative multi-agent systems and serve as ideal models for multi-agent reinforcement learning in frontier artificial intelligence applications, ranging from complex board games to planning for intelligent and autonomous systems \citep{ref:Zhang21}. Consequently, learning in stochastic games has drawn significant interest in examining whether non-equilibrium adaptation of self-interested agents converges to equilibrium, particularly in zero-sum or identical-interest settings \citep{ref:Baudin22,ref:Leslie20,ref:Sayin20,ref:Sayin21,ref:Sayin22}. These findings enhance the predictive power of equilibrium analysis \citep{ref:Fudenberg09}. However, for control and optimization, a drawback is that convergence guarantees apply only to certain equilibria, and agents using these dynamics may perform arbitrarily poorly even in stochastic teams with a common goal. For instance, while efficient learning dynamics have been extensively studied in repeated games \citep{ref:Bistriz21,ref:Chasparis13,ref:Marden12,ref:Marden14,ref:Pradelski12}, results for stochastic games remain limited \citep{ref:Yongacoglu22}.

    \subsection{Contributions}
    In this paper, we provide an affirmative answer to the question \textit{whether we can address efficient learning in stochastic teams through non-equilibrium adaptation of self-interested agents.} {  We present a \textit{new family} of logit-Q dynamics combining the classical and independent log-linear learning dynamics \citep{ref:Marden12,ref:Tatarenko17} with Q-learning \citep{ref:Watkins92} within the \textit{stage-game framework}. The family members differ in terms of when agents revise their actions and how they model their joint play, as tabulated later in \Cref{tab:family}.} In the stage-game framework, we view stochastic games as agents are repeatedly playing some stage game associated with the current state of the underlying stochastic game while the payoffs of these stage games are determined by the agents' immediate and discounted expected continuation payoffs, called \textit{Q-function}. 
    
    The stage-game framework originates with stochastic games \citep{ref:Shapley53} and is used in Minimax-Q \citep{ref:Littman94} and Nash-Q \citep{ref:Hu03} to compute equilibria. Recently, it has been employed by \cite{ref:Baudin22,ref:Leslie20,ref:Sayin20,ref:Sayin21,ref:Sayin22} to blend best response and fictitious play with value iteration for learning in stochastic games. Following this trend, in the logit-Q dynamics, we let agents follow { log-linear learning in the stage games specific to each state whenever the associated state gets visited as if the stage-game payoffs are their Q-function estimates so that they can learn to coordinate in the soft maximizers of the stage games. Without knowing the underlying dynamics, the agents learn Q-functions in a model-free manner akin to (soft) Q-learning \citep{ref:Watkins92} as if their estimated play is the (soft) maximizer. We consider two schemes for how agents can model their past play: In the \textit{averaging} approach, agents take the empirical average of the actions played. In the \textit{exploration-free} approach, agents use the most frequently played actions since they play the best response with higher probability in the log-linear learning dynamics.}
    
    We show that in the logit-Q dynamics, the Q-function estimates converge to the Q-functions associated with the efficient equilibrium and the { estimated play reaches} efficient equilibrium in stochastic teams approximately almost surely, with quantifiable approximation errors (See \Cref{thm:main}). Furthermore, we show the rationality of the dynamics against agents following pure stationary strategies (See \Cref{cor:rational}).

    Our framework extends beyond stochastic teams to accommodate a broader class of potential games. Specifically, consider scenarios where agents' stage payoffs align with common potential functions at each state. While these stage payoffs induce potential games, the continuation payoffs may not necessarily align with the potential function derived from these payoffs. As a result, the overall stochastic game may fail to be a potential game \citep{ref:Leonardos21}. However, in single-controller stochastic games—where a single agent controls state transitions—we can establish that the stage games remain potential games under such stage payoffs, regardless of the continuation payoffs \citep{ref:Sayin22}. Consequently, in single-controller stochastic games, logit-Q dynamics still converge to an equilibrium efficient for the single controller, provided that the single controller's stage payoffs are consistent with the potential function induced by all agents' stage payoffs (See \Cref{cor:potential}).
    \color{black}
    
    \subsection{Challenges}
    A stochastic game is a multi-agent generalization of Markov decision processes (MDPs) in which agents jointly control the \textit{immediate payoff} they receive and the evolution of the underlying state and, therefore, the \textit{continuation payoff}. The trade-off between immediate and continuation payoffs poses a challenge for the generalization of the existing results in efficient learning for games with repeated play (such as \citep{ref:Chasparis13,ref:Marden12,ref:Marden14,ref:Pradelski12}) to stochastic games. The immediate payoff function is stationary, whereas the continuation payoff function is not necessarily stationary by depending on the agents' evolving strategies. Recently, the non-stationarity challenge for learning in stochastic games has been addressed via the \textit{two-timescale learning framework} for the best response and fictitious play dynamics \citep{ref:Leslie20,ref:Sayin20,ref:Sayin22}. However, these dynamics have convergence guarantees only for some  equilibrium, possibly inefficient with respect to the common team objective. 
    
    The log-linear learning update can attain the efficient equilibrium in the repeated play of identical-interest (or potential) games in the sense that the agents play the pure efficient equilibrium at high percentage of time in the long run \citep{ref:Marden12}. Recall the challenge that the stage-game payoffs, i.e., Q-function estimates, are not necessarily stationary. Therefore, the key question is whether the action profiles played by the agents can track the efficient equilibrium associated with the evolving Q-function estimates of the agents. The two-timescale stochastic approximation methods applied in \citep{ref:Leslie20,ref:Sayin20,ref:Sayin21,ref:Sayin22} are not applicable to this problem as the log-linear learning update is not in the form of a stochastic approximation recursion different from the fictitious play and best-response dynamics. 
    
    We develop a new approach to address the non-stationarity challenge for the logit-Q dynamics. Particularly, we divide the horizon into epochs {  with growing lengths} only for analysis so that we can bound how much the Q-function estimates can change and the vanishing step sizes used in the Q-function updates imply that these Q-function estimates, and therefore, the stage games become quasi-stationary within these epochs. To address quasi-stationarity, we consider a fictional scenario where the Q-function estimates do not change so that we have stationary stage games within these epochs. For the fictional scenario, the logit-Q dynamics reduce to the classical or independent log-linear learning dynamics for the repeated play of each stage game and the action profiles played by the agents form an irreducible and aperiodic Markov chain. Correspondingly, we can view the change in the Q-function estimates as perturbations for the Markov chain in the fictional scenario. However, the perturbation is not necessarily stationary or convergent, and the existing perturbation bounds for the analysis of Markov chains (e.g., see \citep{ref:Cho01} for an overview) are not tight enough for our analysis. Therefore, we formulate a new perturbation bound (see \Cref{lem:coupling}) to approximate the game dynamics with its unperturbed version by coupling the evolution of the dynamics in the perturbed and unperturbed versions. \Cref{lem:coupling} can be of independent interest.
    
    \subsection{Related Works} 
    Stochastic teams have drawn significant attention in multi-agent learning. Early studies focused on computing Q-function estimates for efficient equilibria in stochastic teams with unknown model \citep{ref:Boutilier96,ref:Lauer00,ref:Littman01,ref:Wang02}. However, these approaches are based on the maximization over action profiles in a coordinated manner rather than agents are learning to coordinate in the maximizing joint actions in an uncoupled way without taking the objective of other agents into account, as in the logit-Q dynamics for stochastic teams or as in \citep{ref:Chasparis13,ref:Marden12,ref:Marden14,ref:Pradelski12} for the repeated play of games. Such uncoupled learning dynamics are of interest for \textit{non-cooperative} environments (e.g., non-cooperative coordination problems) as they are consistent with the self-interest of the agents to a certain extent and can lead to rational response against opponents playing according to some stationary strategies. 
    
    For example, in the classical log-linear learning, $(i)$ agents believe that the others would play their latest actions (as in Cournot learning, e.g., see \citep{ref:Cheung97}), $(ii)$ the agents respond according to the smoothed best response to that belief when they update their actions, and $(iii)$ the agents take random turns in updating their actions. Experimental studies suggest that human behavior in games is closer to Cournot learning than the widely studied fictitious play \citep{ref:Cheung97}. In \citep{ref:Marden12}, the authors showed the efficiency of the classical log-linear learning and its independent and payoff-based variants in achieving the optimal equilibrium in potential games (and more generally weakly acyclic games) played repeatedly. Later, in \citep{ref:Marden14} and \citep{ref:Pradelski12}, the authors addressed efficient learning in general class of games beyond potential or weakly acyclic games by letting agents update (or not) depending on whether they are in the content or discontent states according to \textit{benchmark} payoffs and actions they keep track of. In \citep{ref:Chasparis13}, the authors presented a learning dynamic where agents continue to play an action as long as the received payoff exceeds a specified aspiration level in coordination games. The common theme in the dynamics presented in these works \citep{ref:Chasparis13,ref:Marden12,ref:Marden14,ref:Pradelski12} are recency in the beliefs formed, and exploration and friction in action revision to selectively attain the efficient outcome. Furthermore, these results are only for the repeated play of games. On the other hand, recently, fictitious play and policy gradient based dynamics have been presented for learning in stochastic teams, e.g., see \citep{ref:Baudin22,ref:Sayin22,ref:Leonardos21}; however, they provide convergence guarantees only for some (possibly inefficient) equilibrium.
    
    Recent studies on logit-Q dynamics include \citep{ref:Bistriz21}, \citep{ref:Yongacoglu22}, and \citep{ref:Onur23}. In \citep{ref:Bistriz21}, efficient learning is examined for stochastic games with exogenous state transitions. Hence, the agents do not face the trade-off between immediate and continuation payoffs, whereas this trade-off is the core challenge for learning in general classes of stochastic games.
    In \citep{ref:Yongacoglu22}, the authors focused on efficient learning in stochastic games by restricting the strategy spaces of the agents to pure stationary strategies and by dividing the horizon into phases within which the agents do not update their strategies so that they can learn the associated Q-functions based on local information only. Lastly, in \citep{ref:Onur23}, a precursor of this paper, we focused on \textit{episodic} logit-Q dynamics, where agents do not update their Q-function estimates within the episodes, and therefore, the stage games they play are stationary (rather than quasi-stationary). However, such episodic approaches followed in \citep{ref:Yongacoglu22} and  \citep{ref:Onur23} require the coordination of the agents in not updating certain parameters for certain time intervals periodically. 
    
    \subsection{Organization} 
    The paper is organized as follows. We describe stochastic games in \Cref{sec:model} and the logit-Q dynamics in \Cref{sec:dynamics}. We present the convergence result and the proofs, resp., in \Cref{sec:result,sec:proofmain}. {In \Cref{sec:examples}, we provide numerical examples, corroborating the convergence results in \Cref{sec:result}.} We conclude the paper with some remarks in \Cref{sec:conclusion}.
    
    \section{Stochastic Games}\label{sec:model}
    Formally, an $N$-agent stochastic game, introduced by \cite{ref:Shapley53}, is a multi-stage game played over infinite horizon that can be characterized by the tuple $\langle S,(A^i)_{i\in[N]},(r^i)_{i\in [N]},p,\gamma\rangle$. At each stage $t=0,1,\ldots$, the game visits a state $s$ from a \textit{finite} set $S$ and each agent $i$ simultaneously takes an action $a^i$ from a \textit{finite} set $A^i$ to collect \textit{stage-payoffs}. The formulation can be extended to state-variant action sets rather straightforwardly. Agent $i$'s stage-payoff is denoted by $r^i:S\times A \rightarrow [0,1]$. Similar to MDPs, stage-payoffs and transition of the game across states depend only on the current state visited and current \textit{action profile} $a=\{a^i\}_{i\in [N]}$ played. Particularly, $p(s' \mid s,a)$ for each $(s,a,s')\in S\times A\times S$ denotes the probability of transition from $s$ to $s'$ under action profile $a$, where $A:=\prod_{i}A^i$. Lastly, $\gamma\in [0,1)$ is the discount factor.
    
    Assume that the agents can correlate their randomized actions and consider joint Markov stationary strategies $\pi:S\rightarrow \Delta(A)$ determining the probability of the joint actions to be played by all agents contingent on the current state.\footnote{We let $\Delta(A)$ denote the probability simplex over the finite set $A$.} Given $\pi$ and the \textit{initial} state $s\in S$, agent $i$'s utility is given by
    \be\label{eq:utility}
    U^i(\pi;s) := \E_{a_t\sim\pi(s_t)}\left[\sum_{t=0}^{\infty}\gamma^tr^i(\bs{s}_{t}^{},\bs{a}_{t}) \mid s_0 = s\right],
    \ee
    where $(s_t,a_t)$ denotes the pair of state and action profile realized at stage $t$. The expectation is taken with respect to the randomness on $(s_t,a_t)$ induced by the underlying transition kernel and the joint strategy $\pi$.

    Given a joint Markov stationary strategy $\pi:S\rightarrow \Delta(A)$, we denote the joint strategy of all agents except $i$ as $\pi^{-i}$, where $\pi^{-i}(s)\in \Delta(A^{-i})$ represents the marginal distribution of $\pi(s)$ over $A^{-i}:= \prod_{j\neq i}A^j$. Additionally, we define the joint strategy that randomizes actions at each state $s$ using the product distribution $\tilde{\pi}^i(s)\times \pi^{-i}(s) \in \Delta(A)$ as $\tilde{\pi}^{i}\times \pi^{-i}$. Below, we introduce equilibrium concepts for such joint Markov strategies.
    
    \begin{definition}\label{def:CCE}
    A Markov stationary strategy $\pi:S\rightarrow \Delta(A)$ is an $\epsilon$-approximate perfect Markov coarse correlated equilibrium (abbreviated as $\epsilon$-CCE) if it satisfies 
    \be\label{eq:CCE}
    U^i(\pi;s) \geq U^i(\tilde{\pi}^i\times \pi^{-i};s) - \epsilon
    \ee 
    for all $\tilde{\pi}^i:S\rightarrow \Delta(A^i)$, $i\in [N]$, and $s\in S$ \citep[Definition 2.1]{ref:Daskalakis23}, where $\epsilon\geq0$.
    \end{definition}
    
    Notice that CCEs are defined for joint Markov stationary strategies. If we restrict our attention to product strategies, we recover the definition of Nash equilibria: an $\epsilon$-CCE becomes an $\epsilon$-approximate perfect Markov Nash equilibrium if the joint strategy is a product strategy, i.e., $\pi=\bigtimes_i \pi^i$ for some $\pi^i:S\rightarrow\Delta(A^i)$ and $i\in [N]$ \citep[Definition 2.2]{ref:Daskalakis23}. 
    
    In stochastic games, a perfect Markov Nash equilibrium (also known as Markov perfect equilibrium) always exists \citep{ref:Fink64}, implying the existence of CCE as well. However, multiple equilibria may arise, leading to different outcomes in terms of performance metrics. In identical-interest stochastic games (also known as stochastic teams), where all agents share the same reward function, i.e., $r^i(s,a)=r(s,a)$ for all $(s,a)$ and some common $r:S\times A\rightarrow\mathbb{R}$, a natural performance metric is the common utility, defined as
    \be\label{eq:team_obj}
    U(\pi;s) := \E_{a_t\sim\pi(s_t)}\left[\sum_{t=0}^{\infty}\gamma^tr(\bs{s}_{t}^{},\bs{a}_{t}) \mid s_0 = s\right].
    \ee 
    Correspondingly, we say that a joint Markov stationary strategy $\pi:S\rightarrow\Delta(A)$ is $\epsilon$-efficient for some $\epsilon>0$ if $U(\pi;s) \geq U(\tilde{\pi};s) - \epsilon$ for all $\tilde{\pi}:S\rightarrow \Delta(A)$ and $s\in S$. By definition, an $\epsilon$-efficient strategy is also an $\epsilon$-CCE. Moreover, an efficient joint strategy always exists, as it corresponds to the unique solution of the Bellman optimality equations for $U$ which are guaranteed to exist \cite{ref:Bertsekas17}.
    
    \color{black}

    In this paper, we present learning dynamics reaching such efficient equilibria. To this end, we focus on the \textit{stage-game framework}, going back to the introduction of stochastic games by \cite{ref:Shapley53}. Particularly, given the joint Markov stationary strategy $\pi$, let $v_{\pi}^i:S\rightarrow\mathbb{R}$ be the agent $i$'s \textit{value function} such that $v^i_{\pi}(s)$ is the value of state $s$ for agent $i$ if the game starts at state $s$ and the agents play according to $\pi$, i.e., $v^i_{\pi}(s) = U^i(\pi;s)$. Correspondingly, let $Q^i_{\pi}:S\times A \rightarrow \mathbb{R}$ be the agent $i$'s \textit{Q-function} such that $Q^i_{\pi}(s,a)$ is the value of state-action pair $(s,a)$ for agent $i$. By \eqref{eq:utility}, the Bellman operator for a strategy yields that
    \begin{subequations}\label{eq:local}
    \begin{flalign}
    &Q^i_{\pi}(s,a) = r^i(s,a) + \gamma \sum_{s'\in S} p(s'\mid s,a)\cdot v^i_{\pi}(s')\quad\forall (s,a),\label{eq:Qlocal}\\
    &v^i_{\pi}(s) = \E_{a\sim \pi(s)}\left[Q^i_{\pi}(s,a)\right]\quad\forall s.\label{eq:valuelocal}
    \end{flalign}
    \end{subequations}
    Then, we can view stochastic games as the repeated play of stage games whose payoffs are the Q-functions associated with how they play in these stage games, whenever the associated state gets visited. 
    
    \section{Logit-Q Dynamics for Stochastic Games}\label{sec:dynamics}
    
    In this section, we introduce the \textit{logit-Q dynamics} within the stage-game framework. In this framework, agents follow \textit{(independent) log-linear learning} in each stage game as if the stage-game payoffs are the Q-functions estimated. They estimate the Q-functions associated with how they play in these stage games by solving the fixed-point problem \eqref{eq:local} iteratively based on the game history.
    To this end, each agent $i$ keeps track of
    \begin{itemize}
    \item $a_t^j(s)\in A^j$ for each $j\in [N]$ denoting agent $j$'s latest action at state $s$ until and including stage $t$, with a slight abuse of notation,
    \item $c_t(s,a)$ and $c_t(s)\in \{0,1,\ldots\}$ denoting, resp., the number of times the pair $(s,a)$ and $s$ get realized until and including stage $t$,
    \item $Q_t^i(s,a)\in \mathbb{R}$ denoting the Q-function estimate of the pair $(s,a)$ at stage $t$.
    \end{itemize}
    
    Recall the friction in the action updates of the log-linear learning, also known as the \textit{lock-in property}. To model such friction, we consider an i.i.d. process $\{I_t\subset [N]\}_{t\geq 0}$ determining the agents updating their actions. For example, agent $i$ revises his/her action at stage $t$ if and only if $i \in I_t$. For the distribution of the revision process, we consider two schemes. In the classical \textit{coordinated} scheme, we have $I_t\sim \mathcal{P}^{\mathrm{coor}}$ and\footnote{For a finite set $I$, we denote its number of elements by $|I|$.} only a randomly selected agent can update his/her action. In the \textit{independent} scheme, we have $I_t\sim \mathcal{P}^{\mathrm{ind}}$ and agents update their actions independently with probability $\delta\in (0,1)$. The distributions $\mathcal{P}^{\mathrm{coor}}$ and $\mathcal{P}^{\mathrm{ind}}$ are defined as follows:
    \be\label{eq:Pcoor}
    \mathcal{P}^{\mathrm{coor}}(I) := \mathds{1}_{\{|I|=1\}}\frac{1}{N}\quad\forall I\subset [N], \quad \mathcal{P}^{\mathrm{ind}}(I) = \delta^{|I|}(1-\delta)^{N-|I|}\quad\forall I\subset [N],
    \ee
    
    Let $s_t\in S$ be the state at stage $t$. Each agent $i\in I_t$ revises his/her action according to the \textit{smoothed} best response \textit{as if} the game payoffs are $Q_t^i(s_t,\cdot)$ and the other agents are playing the latest actions $a_{t-1}^{-i}(s_t) = \{a_{t-1}^j(s_t)\}_{j\neq i}$. Particularly, given $q\in \mathbb{R}^{B}$ for some finite set $B$, we define the smoothed best response by
    \be\label{eq:logit}
    \br(q)(b) := \frac{\exp(q(b)/\tau)}{\sum_{b'\in B}\exp(q(b')/\tau)}\quad\forall b\in B.
    \ee
    Then, we have $a_t^i\sim \br(Q_t^i(s_t,\cdot,a_{t-1}^{-i}(s_t)))$ and the parameter $\tau>0$ balances exploration vs exploitation. For example, agent $i$ plays best response(s) with higher probabilities (whose sum goes to $1$ as $\tau\rightarrow 0$) and can explore any action with some positive probabilities (going to $1/|A^i|$ as $\tau\rightarrow \infty$). On the other hand, each agent $j\notin I_t$, not revising their actions, plays their latest action at that state, i.e., $a_t^j = a_{t-1}^j(s_t)$. 
    
    Agents estimate the joint Markov stationary strategy \(\pi:S\to\Delta(A)\) from the game history in one of two ways. In the \emph{averaging} approach, the estimated play is the average of the joint actions, and in the \emph{exploration-free} approach, which mitigates exploration effects (due to \(\tau>0\) in \eqref{eq:logit}), the estimated play is the most frequently played actions. For both cases, we define $\pi_t^{\mathrm{ave}}$, and $\pi_t^{\mathrm{free}}$ respectively as follows:
    \begin{flalign} \label{eq:piave}
    \pi_t^{\mathrm{ave}}(s)(a)=\frac{c_t(s,a)}{c_t(s)} \quad, \quad \pi_t^{\mathrm{free}}(s)(a)=
    \begin{cases}
    1/|A_t(s)| & \text{if } a\in A_t(s),\\[1mm]
    0 & \text{otherwise},
    \end{cases} \quad\forall (s,a),
    \end{flalign}
    with $A_t(s)=\arg\max_a\{c_t(s,a)\}\subset A$.
    
    Let $\pi_t^{\mathrm{ave}}(s)$ and $\pi_t^{\mathrm{free}}(s)$ be uniform for $c_t(s)=0$, without loss of generality.
    
    To compute the Q-function associated with the estimated play, the agents can solve \eqref{eq:local} iteratively and use \textit{one-stage look ahead} to address the unknown transition kernel, as in the classical Q-learning. After observing the next state $s_{t+1}$ at stage $t+1$, each agent $i$ updates his/her Q-function estimate for the pair $(s_t,a_t)$ according to
    \be
    \label{eq:Q}
    Q_{t+1}^i(s_t,a_t)= Q_t^i(s_t,a_t) + \beta_{c_t(s_t,a_t)} \left(r^i_t + \gamma v_t^i(s_{t+1})  - Q_t^i(s_t,a_t)\right),
    \ee
    where $r_t^i:= r^i(s_t,a_t)$ is the stage-payoff received and $v_t^i$ denotes the value function estimate. Here, the agents use the value $v_t^i(s_{t+1})$ of the next state as an approximation of the expected continuation payoff $\sum p(s'\mid s,a)  v_t^i(s')$ in \eqref{eq:Qlocal}. Given the estimated play $\pi_t$, the value $v_t^i$ is given by
    \be
    v_t^i(s_{t+1}) = \E_{a\sim \pi_t(s_{t+1})}[Q_t^i(s_{t+1},a)]
    \ee
    due to \eqref{eq:valuelocal}. The \textit{reference} step size $\beta_c\in (0,1)$ for all $c=0,1,\ldots$ implies that the agents update their Q-function estimates to convex combinations of the previous estimates $Q_t^i$ and the stage-games' (estimated) payoffs $r^i_{t} + \gamma \E_{a\sim \pi_{t}(s_{t+1})}[Q_{t}^i(s_{t+1},a)]$. Since agents do not observe the counterfactuals about the joint actions that are not played and the states that are not visited in the current stage, they do not update the other Q-function estimates, i.e., $Q_{t+1}^i(s,a) = Q_{t}^i(s,a)$ for all $(s,a)\neq (s_t,a_t)$. 
    
    \begin{algorithm}[tb]
    \caption{Logit-Q Dynamics}
    \begin{algorithmic}[1]\label{alg}
    \REQUIRE each agent can observe $(s_t,a_t)$ and keep track of $a_t(\cdot)$ and $c_{t}(\cdot)$
    \STATE {\bfseries Input:} initializations $Q^i_{0}(s, a)=0$ for all $(i,s,a) \in [N]\times S \times A$.
    \FOR{each stage $t=0,1,\ldots$}
    \STATE $s_t$ gets realized
    \STATE {\bfseries if} $t=0$ {\bfseries then} jump to Step \ref{step:play}
    \STATE each agent $i$ {\bfseries updates} the Q-function estimates for stage $t-1$ according to
    \[
    Q_{t}^i(s,a)= Q_{t-1}^i(s,a) + \mathds{1}_{\{(s,a)=(s_{t-1},a_{t-1})\}}\cdot\beta_{c_{t-1}(s,a)} \cdot(\what{Q}_{t}^i  - Q_{t-1}^i(s,a)),
    \]
    where $\what{Q}_t^i := r^i_{t-1} + \gamma \E_{a\sim \pi_{t-1}(s_t)}[Q_{t-1}^i(s_{t},a)]$ and $\pi_{t-1}(\cdot)$ depends on $c_{t-1}(\cdot)$
    \STATE $I_t\sim \mathcal{P}$ gets realized \label{step:play}
    \STATE each agent $i\in I_t$ {\bfseries plays} $a^i_t \sim \br(Q_t^i(s_t,\cdot,a_{t-1}^{-i}(s_t)))$ 
    \STATE each agent $i\notin I_t$ {\bfseries plays} $a^i_t = a_{t-1}^i(s_t)$
    \ENDFOR
    \end{algorithmic}
    \end{algorithm}
    
    \begin{table}[t!]
    \centering
    \caption{A family of the logit-Q dynamics with the parameters $\mathcal{P}$ and $\pi_t$, as described in \eqref{eq:Pcoor}, and \eqref{eq:piave}}\label{tab:family}
    \begin{tabular}{l|m{.6in}|m{.6in}}
    Logit-Q Dynamics & Revision Process $\mathcal{P}$ & Estimated Play $\pi_t$\\
    \hline\hline
    Averaged Logit-Q (AL-Q) & $\mathcal{P}^{\mathrm{coor}}$ & $\pi_t^{\mathrm{ave}}$\\
    \hline
    Exploration-free Logit-Q (EL-Q) & $\mathcal{P}^{\mathrm{coor}}$ & $\pi_t^{\mathrm{free}}$\\
    \hline
    Averaged Independent Logit-Q (AIL-Q) & $\mathcal{P}^{\mathrm{ind}}$ &$\pi_t^{\mathrm{ave}}$\\
    \hline
    Exploration-free Independent Logit-Q (EIL-Q) & $\mathcal{P}^{\mathrm{ind}}$ & $\pi_t^{\mathrm{free}}$
    \end{tabular}
    \end{table}
    
    \Cref{alg} describes a family of logit-Q dynamics that differ in their revision processes $\mathcal{P}$ and estimated plays $\pi_t$, as tabulated in \Cref{tab:family}. Recall that agents update their Q-functions after observing the next state and they initialize Q-function estimates as $Q_0^i(s,a)=0$ for all $(i,s,a)$ rather than arbitrarily. While such coordination can be challenging for ad hoc team formations, the step sizes $\beta_c\in(0,1)$ suppress the impact of the initialization on the estimates at the future stages and we examine the impact of arbitrary initializations numerically later in \Cref{sec:examples}. Moreover, this initialization and $\beta_c\in(0,1)$ ensure that $0\leq Q_t^i(s, a)\leq 1/(1-\gamma)$ for all $(i,s,a)$ (since $r^i:S\times A\rightarrow [0,1]$). Hence, agents revising their actions always select any action with a probability uniformly bounded away from zero by \eqref{eq:logit}.
    
    \section{Convergence Results}\label{sec:result}
    
    In this section, we characterize the convergence properties of \Cref{alg} based on the following assumption on the step size $\beta_c$. 
    
    \begin{assumption}\label{assume:stepsize}
    The reference step size $\beta_c\in (0,1)$ satisfies
    \begin{enumerate}
    \myitem{$(i)$}\label{subasssume:i} Monotonic decay: $\beta_c\rightarrow 0$ as $c\rightarrow\infty$ and $\beta_c>\beta_{c'}$ for all $c<c'$,
    \myitem{$(ii)$}\label{subasssume:ii} Sufficiently slow decay: $\sum_{c}\beta_c = \infty$,
    \myitem{$(iii)$}\label{subasssume:iii} Sufficiently fast decay: $c\beta_{\lfloor mc \rfloor} \rightarrow 0$ as $c \rightarrow \infty$ for any $m\in (0,1]$.\footnote{Let $\lfloor r\rfloor:=\max\{z\in\mathbb{Z}:z\leq r\}$ for any $r\in \mathbb{R}$ denote the floor function.}
    \end{enumerate} 
    \end{assumption}
    
    \Cref{assume:stepsize}-\ref{subasssume:i} and \Cref{assume:stepsize}-\ref{subasssume:ii} are standard in stochastic approximation, and \Cref{assume:stepsize}-\ref{subasssume:iii} is standard for asynchronous two-timescale stochastic approximation if we view $1/(c+1)$ as the step size for the dynamics evolving at the fast timescale \citep[Section 4.2]{ref:Perkins12}. For example, $\beta_c= 1/((c+1)\log(c+1))$ satisfies \Cref{assume:stepsize}. We also note that \Cref{assume:stepsize}-\ref{subasssume:iii} implies $\sum_{c}\beta_c^2 < \infty$, playing an important role in addressing the stochastic approximation noise. 
    
    The following theorem characterizes the convergence properties of the logit-Q dynamics for stochastic teams.
    
        \begin{table}[]
        \centering
            \caption{Error bounds in \Cref{thm:main}, where $\Lambda(\delta)$ and $\Xi(\delta)$ as described later in \Cref{lem:stationary} and \eqref{eq:defXi}, respectively. Both $\Lambda(\delta)$ and $\Xi(\delta)$ decay to zero as $\delta\rightarrow 0^+$.} \label{tab:bounds}
        \begin{tabular}{c|c|c|c|c}
        & AL-Q & AIL-Q & EL-Q & EIL-Q\\
        \hline\hline
        $e_Q$ \eqref{eq:mainresult} & $\frac{\gamma}{1-\gamma}\cdot (\tau\log|A|)$ & $\frac{\gamma}{1-\gamma}\cdot (\tau\log|A| + \frac{1}{1-\gamma}\cdot\Lambda(\delta))$ & $0$ & $\frac{\gamma}{1-\gamma} \cdot \Xi(\delta)$ 
    \color{black}
        \end{tabular}
        \end{table} 
    

    \begin{theorem}\label{thm:main}
    Given a stochastic team $\langle S,(A^i)_{i\in[N]},r,p,\gamma\rangle$ with the common objective $U(\cdot)$, consider that every agent follow \Cref{alg} with the same revision-process and estimated-play schemes, as in \Cref{tab:family}. Suppose that \Cref{assume:stepsize} holds and the underlying stochastic game is 
    
    irreducible in the sense that there exist at least one Markov stationary strategy, for which the states visited form an irreducible Markov chain.\footnote{Such irreducibility assumptions are widely used for stochastic games, e.g., see \citep{ref:Brafman02,ref:Sayin21}.} 
    \color{black}
    Then, for each $i\in[N]$, we have
    \begin{subequations}\label{eq:result}
    \begin{flalign}
    &\limsup_{t\rightarrow\infty} \left|\,Q_{t}^i(s,a)-Q_*(s,a)\,\right| \leq e_Q\quad\forall (s,a)\label{eq:mainresult}\\
    &\limsup_{t\rightarrow\infty}\left|\max_{\pi}\{U(\pi;s)\}-U(\pi_t;s)\,\right| \leq \frac{1+\gamma}{\gamma(1-\gamma)}\cdot e_Q \quad\forall s\label{eq:resultpi}
    \end{flalign}
    \end{subequations}
    almost surely, where the team-optimal Q-function $Q_*:S\times A\rightarrow\mathbb{R}$ is the unique fixed point of the following Q-function variant of the Bellman optimality operator:
    \begin{align}
    &Q_*(s,a) = r(s,a) + \gamma \sum_{s'} p(s'\mid s,a)\max_{a'\in A} \;\{Q_*(s',a')\}\quad\forall (s,a)\label{eq:Qglobal}
    \end{align}
    and the error bound $e_Q\geq 0$ is as tabulated in \Cref{tab:bounds} for different revision-process and estimated-play schemes.
    \end{theorem}
    
    \Cref{thm:main} shows that in \Cref{alg}, the Q-function estimates and  the estimated plays almost surely converge, resp., to the team-optimal Q-functions and efficient stationary equilibrium \textit{approximately}, with quantifiable error bounds (see \Cref{tab:bounds}). The proof is moved to \Cref{sec:proofmain}.
    
    The following corollary to \Cref{thm:main} shows the rationality of \Cref{alg} against others playing according to pure stationary strategies. 
    
    \begin{corollary}\label{cor:rational}
    Given a general-sum stochastic game $\langle S,(A^i,r^i)_{i\in[N]},p,\gamma\rangle$, consider that agent $i$ follows AIL-Q or EIL-Q while the other agents $j\neq i$ play according to pure Markov stationary strategy, e.g., $\pi^{-i}:S\rightarrow A^{-i}$. Suppose that \Cref{assume:stepsize} holds and the states visited under the strategy $\tilde{\pi}^i\times \pi^{-i}$ form an irreducible Markov chain for any pure Markov stationary strategy $\tilde{\pi}^i:S\rightarrow A^i$. Then, we have
    \begin{subequations}
    \begin{flalign}
    &\limsup_{t\rightarrow\infty} |Q_{t}^i(s,a^i,\pi^{-i}(s))-q_*^i(s,a^i)| \leq e_Q\quad\forall (s,a^i)\label{eq:mainresult2}\\
    &\limsup_{t\rightarrow\infty}|\max_{\tilde{\pi}^i}\{U^i(\tilde{\pi}^i\times \pi^{-i};s)\}-U^i(\pi_t;s)| \leq \frac{1+\gamma}{\gamma(1-\gamma)}\cdot e_Q\quad\forall s\label{eq:resultpi2}
    \end{flalign}
    \end{subequations}
    almost surely, where $q_*^i:S\times A^i\rightarrow\mathbb{R}$ is the local Q-function associated with the best response to $\pi^{-i}$ and is the unique fixed point of the Bellman operator variant
    \be
    q_*^i(s,a^i) = r^i(s,a^i,\pi^{-i}(s)) + \gamma\sum_{s'}p(s'| s,a^i,\pi^{-i}(s)) \max_{\tilde{a}^i\in A^i}\{q_*^i(s',\tilde{a}^i)\}\quad\forall (s,a^i).
    \ee
    The error bound $e_Q\geq 0$ is as in \Cref{tab:bounds} for $A=A^i$.
    \end{corollary}
    
    \begin{proof}
    
    Pure Markov stationary strategy $\pi^{-i}$ yields that the underlying stochastic game reduces to a \textit{single-agent} stochastic team characterized by $\langle S,A^i,\bar{r}^i,\bar{p},\gamma\rangle$ where $\bar{r}^i(s,a^i) = r^i(s,a^i,\pi^{-i}(s))$ and $\bar{p}(s'| s,a^i) = p(s'| s,a^i,\pi^{-i}(s))$. Then, the convergence of logit-Q dynamics in the single-agent stochastic team $\langle S,A^i,\bar{r}^i,\bar{p},\gamma\rangle$ follows from \Cref{thm:main}.
    \color{black}
    \end{proof}

    \subsection{Beyond Stochastic Teams}
    Next, consider an $N$-agent game characterized by $\mathcal{G}=\langle A^i,u^i\rangle_{i\in [N]}$, where $A^i$ is the finite action set of agent $i$, $A:=\prod_{j\in[N]}A^j$ is the set of action profiles, and $u^i:A\rightarrow\mathbb{R}$ is the payoff function of agent $i$. We say that $\mathcal{G}$ is a potential game if there exists a potential function $\Phi:A\rightarrow\mathbb{R}$ satisfying
    \be\label{eq:potential}
    \Phi(a^i,a^{-i}) - \Phi(\tilde{a}^i,a^{-i}) = u^i(a^i,a^{-i}) - u^i(\tilde{a}^i,a^{-i})    
    \ee
    for all $(a^i,a^{-i},\tilde{a}^{i})$ and $i\in [N]$ \citep{ref:Monderer96b}. Note that identical-interest games are potential games with the common payoff as the potential function.

    For potential games, if agents follow (independent) log-linear learning dynamics, then by \eqref{eq:logit} and \eqref{eq:potential}, the smoothed best response can be written as
    \be\label{eq:potupdate}
    \frac{\exp(u^i(a^i,a^{-i})/\tau)}{\sum_{\tilde{a}^i\in A^i}\exp(u^i(\tilde{a}^i,a^{-i})/\tau)} = \frac{\exp(\Phi(a^i,a^{-i})/\tau)}{\sum_{\tilde{a}^i\in A^i}\exp(\Phi(\tilde{a}^i,a^{-i})/\tau)}.
    \ee
    This implies that we can view the (independent) log-linear learning dynamics as if the agents have the potential function as their objective without loss of generality. Therefore, the convergence of the (independent) log-linear learning dynamics to near efficient equilibrium with respect to the potential function follows from the near efficient equilibrium convergence of the dynamics in identical-interest games.

    However, the convergence of \Cref{alg} to near efficient stationary equilibrium in stochastic teams, as shown in \Cref{thm:main}, does not imply the convergence in cases where the stage-payoffs induce a potential game in general. We can list the challenges as follows: 
    
    \textit{Challenge (i).} The stage games may not be potential games even though the stage payoffs induce potential games. By \eqref{eq:potential}, for each state $s$, stage payoffs $r^i(s,\cdot)$'s induce a potential game if there exists a potential function $\Phi(s,\cdot)$ such that
    \begin{equation}\label{eq:potentialbeyond}
    r^i(s,a^i,a^{-i}) - r^i(s,\tilde{a}^i,a^{-i}) = \Phi(s,a^i,a^{-i}) - \Phi(s,\tilde{a}^i,a^{-i}) \quad\forall (\tilde{a}^i,a^i,a^{-i})\mbox{ and }i\in[N].
    \end{equation}
    However, this does not imply that the auxiliary stage game with the payoffs $(Q^i_{\pi}(s,\cdot))_{i\in [N]}$ is a potential game since there may not exist a potential function $\Psi_{\pi}(s,a)$ such that 
    \begin{equation}
    Q_{\pi}^i(s,a^i,a^{-i}) - Q_{\pi}^i(s,\tilde{a}^i,a^{-i}) = \Psi_{\pi}(s,a^i,a^{-i}) - \Psi_{\pi}(s,\tilde{a}^i,a^{-i}) \quad\forall (\tilde{a}^i,a^i,a^{-i})\mbox{ and }i\in[N].
    \end{equation}
    More explicitly, by \eqref{eq:Qlocal} and \eqref{eq:potentialbeyond}, we have
    \begin{flalign}
    &Q_{\pi}^i(s,a^i,a^{-i}) - Q_{\pi}^i(s,\tilde{a}^i,a^{-i}) = r^i(s,a^i,a^{-i}) - r^i(s,\tilde{a}^i,a^{-i}) \nonumber\\
    &\hspace{2.1in} + \gamma \sum_{s_+}p(s_+\mid s,a^i,a^{-i}) v^i_{\pi}(s_+) - \gamma \sum_{s_+}p(s_+\mid s,\tilde{a}^i,a^{-i}) v^i_{\pi}(s_+)\nonumber\\
    &\hspace{1.9in} = \Phi(s,a^i,a^{-i}) - \Phi(s,\tilde{a}^i,a^{-i})\nonumber\\
    &\hspace{2.1in} + \gamma \sum_{s_+}p(s_+\mid s,a^i,a^{-i}) v^i_{\pi}(s_+) - \gamma \sum_{s_+}p(s_+\mid s,\tilde{a}^i,a^{-i}) v^i_{\pi}(s_+)\nonumber\\
    &\hspace{1.9in} = \Psi_{\pi}(s,a^i,a^{-i}) - \Psi_{\pi}(s,\tilde{a}^i,a^{-i}) \quad\forall (\tilde{a}^i,a^i,a^{-i})\mbox{ and }i\in[N].\label{eq:potential2beyond}
    \end{flalign}
    Due to the continuation payoffs $v_{\pi}^i$'s, the existence of a potential function $\Phi$ in the game induced by stage payoffs $r^i$'s is not a sufficient condition for the existence of a potential function $\Psi_{\pi}$ for the auxiliary stage game with payoffs $Q_{\pi}^i$'s.

    \textit{Challenge (ii).} Even when stage games are potential games, reaching the efficient equilibrium in the stage games with respect to the potential function does not imply the maximization of the local stage-game payoffs. Observe that maximization of the potential function $\Psi_{\pi}(s,\cdot)$ does not necessarily imply the maximization of the stage game payoffs $Q_{\pi}^j(s,\cdot)$'s since there is no general guarantee that 
    \begin{equation}
    \arg\max_{a^j} \Psi_{\pi}(s,a^j,a^{-j}) = \arg\max_{a^j} Q_{\pi}^j(s,a^j,a^{-j}).
    \end{equation} 

    The following corollary to \Cref{thm:main} (similar to \citep[Corollary 4]{ref:Sayin22}) shows that we can address these challenges for a certain class of stochastic games.

    \begin{corollary}\label{cor:potential}
    Consider a general-sum stochastic game $\langle S,(A^j,r^j)_{j\in[N]},p,\gamma\rangle$ where for some specific agent $i$, the stage-payoffs of agents $j\neq i$ satisfy
    \be\label{eq:potential2}
    r^i(s,a^j,a^{-j}) - r^i(s,\tilde{a}^j,a^{-j}) = r^j(s,a^j,a^{-j}) - r^j(s,\tilde{a}^j,a^{-j}) \quad\forall (s,a^j,a^{-j},\tilde{a}^j)
    \ee
    similar to \eqref{eq:potential}, and the underlying transition kernel satisfies
    \be\label{eq:single}
    p(s'\mid s,a^i,a^{-i}) =  p(s'\mid s,a^i,\tilde{a}^{-i})\quad\forall (s,a^i,a^{-i},\tilde{a}^{-i},s')
    \ee
    such that only agent $i$ controls the state transitions.

    Suppose that every agent follow \Cref{alg}, \Cref{assume:stepsize} holds and the underlying stochastic game is irreducible, as described in \Cref{thm:main}. Then, for the single-controller agent $i$, we have
    \begin{subequations}
    \begin{flalign}
    &\limsup_{t\rightarrow\infty} \left|\;Q_{t}^i(s,a)-Q_*^i(s,a)\;\right| \leq e_Q \quad\forall (s,a)\label{eq:mainresult3}\\
    &\limsup_{t\rightarrow\infty}\left|\;\max_{\pi}\{U^i(\pi;s)\} - U^i(\pi_t;s) \;\right| \leq  \frac{1+\gamma}{\gamma(1-\gamma)}\cdot e_Q\quad\forall s\label{eq:resultpi3}
    \end{flalign}
    \end{subequations}
    almost surely, where the Q-function $Q_*^i:S\times A\rightarrow\mathbb{R}$ is the unique fixed point of the following Bellman operator variant:
    \be
    Q_*^i(s,a) = r^i(s,a) + \gamma \sum_{s'} p(s'\mid s,a)\max_{a'\in A} \;\{Q_*^i(s',a')\}\quad\forall (s,a).\label{eq:Qiglobal}
    \ee
    The error bounds are as described in \Cref{tab:bounds}.
    \end{corollary}

    \begin{proof}
    By \eqref{eq:single}, we define $\bar{p}(s_+\mid s,a^i) := p(s_+\mid s,a^i,a^{-i})$ and $\Psi_{\pi}(s,a^i,a^{-i}) := Q_{\pi}^i(s,a^i,a^{-i})$ for all $(s,a^i,a^{-i},s_+)$. Note that $i$ refers to the single controller rather than any agent. Then, \eqref{eq:Qlocal} yields that
    \begin{equation}\label{eq:Psi}
    \Psi_{\pi}(s,a^i,a^{-i}) := r^i(s,a^i,a^{-i}) + \gamma \sum_{s' \in S}\bar{p}(s'\mid s,a^i) \cdot v^i_{\pi}(s').
    \end{equation}
    By the definition $\Psi_{\pi}\equiv Q_{\pi}^i$, $\Psi_{\pi}$ is aligned with $Q_{\pi}^i$ and \eqref{eq:potential2beyond} holds for the single controller. For agents $j\neq i$, we have
    \begin{flalign}
    &Q_{\pi}^j(s,a^j,a^{-j}) - Q_{\pi}^j(s,\tilde{a}^j,a^{-j}) = r^j(s,a^j,a^{-j}) - r^j(s,\tilde{a}^j,a^{-j}) \nonumber\\
    &\hspace{2.1in} + \gamma \sum_{s'}\bar{p}(s'\mid s,a^i) v^j_{\pi}(s') - \gamma \sum_{s'}\bar{p}(s'\mid s,a^i) v^j_{\pi}(s')\label{eq:rjs}\\
    &\hspace{1.9in} \stackrel{(a)}{=} r^i(s,a^j,a^{-j}) - r^i(s,\tilde{a}^j,a^{-j})\\
    &\hspace{1.9in} \stackrel{(b)}{=} r^i(s,a^j,a^{-j}) - r^i(s,\tilde{a}^j,a^{-j})\nonumber\\
    &\hspace{2.1in} + \gamma \sum_{s'}\bar{p}(s'\mid s,a^i) v^i_{\pi}(s') - \gamma \sum_{s'}\bar{p}(s'\mid s,a^i) v^i_{\pi}(s')\\
    &\hspace{1.9in} \stackrel{(c)}{=} \Psi_{\pi}(s,a^i,a^{-i}) - \Psi_{\pi}(s,\tilde{a}^i,a^{-i}) \quad\forall (\tilde{a}^i,a^i,a^{-i})\mbox{ and }i\in[N],
    \end{flalign}
    where $(a)$ follows from \eqref{eq:potential2} and since the last two terms on the right-hand side of \eqref{eq:rjs} cancel each other, $(b)$ follows from adding and subtracting $\gamma \sum_{s'}\bar{p}(s'\mid s,a^i) v^i_{\pi}(s')$, and $(c)$ is due to \eqref{eq:Psi}. Therefore, \eqref{eq:potential2beyond} holds also for agents $j\neq i$. This implies that the auxiliary stage games with the payoffs $(Q_{\pi}^j)_{j\in [N]}$ are potential games with the potential functions $\Psi_{\pi}$. Furthermore, the maximization of the potential function  $\Psi_{\pi}$ in the stage games would imply the maximization of the stage game payoffs $Q_{\pi}^i$ of the single controller $i$ since $\Psi_{\pi}\equiv Q_{\pi}^i$. Due to \eqref{eq:potupdate}, the logit-Q dynamics for the stochastic game $\langle S,(A^j,r^j)_{j\in[N]},p,\gamma\rangle$ would lead to the same logit-Q dynamics in the stochastic team $\langle S,(A^j)_{j\in [N]},r^i,p,\gamma\rangle$. Then, the proof follows from \Cref{thm:main}.  
    \end{proof}
    
    \section{Proof of \Cref{thm:main}}\label{sec:proofmain}
    The proof follows by showing that the estimated plays $\pi_t$ track the (soft) maximizers of the Q-function estimates $Q_t^i$'s. Then, the Q-update \eqref{eq:Q} approximates the (soft) Q-learning algorithm in which the entire team coordinates on the (soft) maximizer, and under standard conditions on step sizes and transition probabilities, Q-learning converges to the Q-function associated with the optimal team strategy.
    
    In the following, we show this \textit{tracking result} by focusing on the evolution of the estimated plays across the repeated play of stage games. Recall that in the logit-Q dynamics, agents follow the \textit{logit dynamics} in the stage games specific to each state whenever the associated state gets visited as if the stage-game payoffs are their Q-function estimates. Hence, we first characterize the convergence properties of the logit dynamics for non-stationary environments, which might be of independent interest.
    
    \subsection{Convergence Properties of the Logit Dynamics}
    
    In the logit dynamics, agents revise their actions according to a revision process, e.g., either in a coordinated or independent way. The agents revising their actions play according to the smoothed best response to the others' previous actions. 
    
    Notably, the action profiles played in the logit dynamics form an irreducible and aperiodic Markov chain \citep{ref:Marden12}. The following proposition characterizes the dependence of the (long-run) behavior of this Markov chain on the game payoffs. 
    
    \begin{proposition} \label{prop:propConstantPayoff}
    Consider that agents follow the logit dynamics in the repeated play of a potential game $\mathcal{G}=\langle (A^i,u^i)_{i\in [N]}\rangle$ with the potential function $\Phi$. Then, the transition matrix and the unique stationary distribution for the Markov chain formed by the action profiles are locally Lipschitz continuous functions of the potential function. 
    \end{proposition}
    
    \begin{proof}
    In the logit dynamics, the transition probabilities of the Markov chain formed by the action profiles can be written as
    \be\label{eq:transition}
    \Pr(a_+\mid a) = \sum_{I\subset [N]} \mathcal{P}(I) \cdot \prod_{i\in I} \br(\Phi(\cdot,a^{-i}))(a_+^i) \cdot \prod_{j\notin I} \mathds{1}_{\{a_+^j=a^j\}}\quad\forall (a,a_+)\in A\times A,
    \ee
    by noting that $\br(\Phi(\cdot,a^{-i}))=\br(u^i(\cdot,a^{-i}))$ for all $i\in[N]$, where $\mathcal{P}$ is the distribution of the revision process, e.g., as described in \eqref{eq:Pcoor}. Since $\br(\cdot)$ is a continuously differentiable function, \eqref{eq:transition} yields that the transition matrix is a locally Lipschitz continuous function of the potential function. Furthermore, in irreducible and aperiodic Markov chains, the \textit{unique} stationary distributions are Lipschitz continuous functions of their transition matrices \citep[Section 3]{ref:Cho01}.
    \end{proof}
    
    This proposition plays an important role in the convergence analysis since the stage-game payoffs, i.e., Q-function estimates, can evolve in time due to \eqref{eq:Q}. Small changes in the Q-function estimates can result in (relatively) small changes in the transition probabilities and the corresponding stationary distributions due to the Lipschitz continuity.
    
    \begin{figure}
        \centering
        \includegraphics[width=0.9\textwidth]{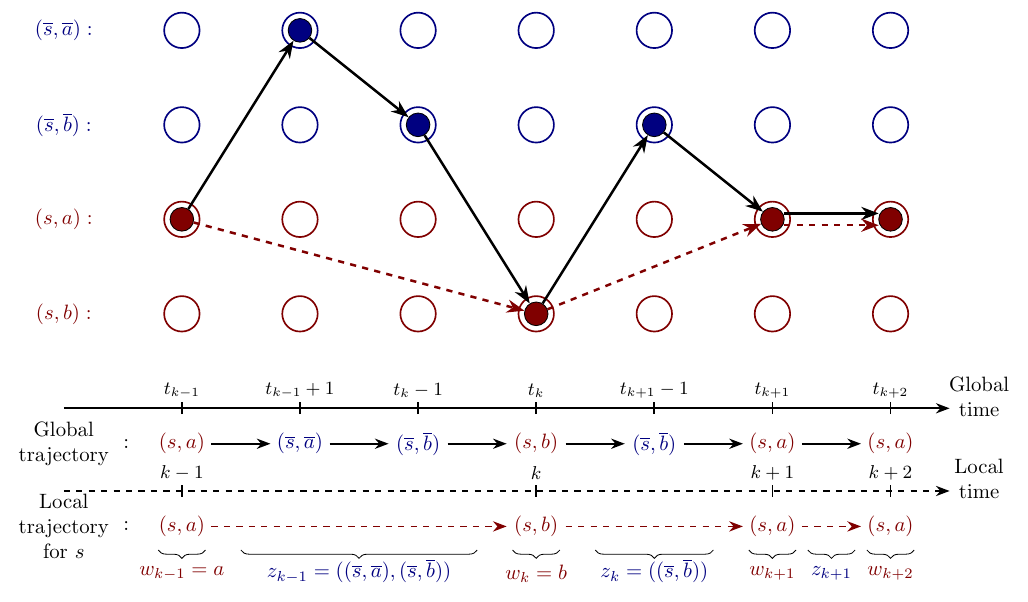}  
        \caption{Consider two states $s$ and $\overline{s}$ with action profiles $\{a,b\}$ and $\{\overline{a},\overline{b}\}$, respectively. This is a figurative illustration for the \textit{global} trajectory of the pairs $(s_t,a_t)_{t\geq 0}$ evolving over the global timescale $t=0,1,\ldots$, and the \textit{local} trajectory of the pairs $(w_{k},z_{k})_{k\geq 0}$ specific to $s$ and evolving over the local timescale $k=1,2,\ldots$ whenever $s$ gets visited. Let $t_k$ be the time of the $k$th visit to $s$. In the local trajectory, $w_k$ and $z_k$ denote, resp., the action profile played at the $k$th visit to $s$ and the trajectory of the state-action pairs realized in between the $k$th and $(k+1)$th visit to $s$.}\label{fig:processes}
    \end{figure}
    
    In the following, we quantify the impact of these changes on the long-run behavior of the dynamics in comparison to the cases where the underlying stage game is stationary. However, the coupling of the dynamics across states poses a challenge. To address this challenge, we focus on the dynamics specific to a \textit{fixed} state $s$ and model the dynamics specific to other states as some external (possibly coupled) processes. Let $w_k$ and $z_k$ denote, resp., the action profile played at the $k$th visit to $s$ and the trajectory of the state-action pairs realized in between the $k$th and $(k+1)$th visit to $s$, as illustrated in \Cref{fig:processes}. Then, we have $w_k\in A$ and $z_k\in Z$, where $Z$ represents the \textit{countable} set of all possible trajectories in between consecutive visits to $s$. 
    
    Let $t_k$ be the stage for the $k$th visit to $s$. Then, we have $\sigma \left( (s_{\tau}, a_{\tau})_{\tau < t_k}, s_{t_k} \right) = \sigma \left((w_l, z_l)_{l < k} \right)$. The Q-function estimate $Q_{t_k}$ is $\sigma \left((w_l, z_l)_{l < k} \right)$-measurable since $Q_{t_k}$ is $\sigma \left((s_{\tau}, a_{\tau})_{\tau < t_k}, s_{t_k} \right)$-measurable by \eqref{eq:Q}. Therefore, the randomness on $(w_k,z_k)$ is contingent only on $(w_l, z_l)_{l < k}$ and there exists a transition kernel $\mathbb{P}$ such that $(w_{k}, z_{k}) \sim \mathbb{P} \left( \cdot \mid (w_l, z_l)_{l < k} \right)$. The following lemma can characterize the long-run behavior of the logit dynamics in such non-stationary environments by coupling its evolution with the logit dynamics in stationary environments.
    
    \begin{lemma}\label{lem:coupling}
        Consider two discrete-time processes \( \{ \what{w}_k \}_{k \geq 0} \) and \( \{ (w_k, z_k) \}_{k \geq 0} \) over a finite set \( W \) and the product space \( W \times Z \) with some countable space \( Z \). The processes \( \{ \what{w}_k \} \) and \( \{ (w_k, z_k) \} \) evolve, resp., according to $\what{w}_k \sim \what{\mathbb{P}}(\cdot | \what{w}_{k-1})$ and $(w_k, z_k) \sim \mathbb{P}(\cdot | (w_l, z_l)_{l <k})$. Let $\mathbb{P}^{w}(\cdot | (w_l, z_l)_{l <k})$ denote the marginal of $\mathbb{P}(\cdot | (w_l, z_l)_{l <k})$ over \( Z \) such that $w_k \sim \mathbb{P}^{w}(\cdot | (w_l, z_l)_{l <k})$. Suppose that these processes satisfy the following conditions:
    
    \textbf{Condition $(i)$} There exist $\what{\varepsilon}, \varepsilon > 0$, a destination $w' \in W$ and $\kappa \in \mathbb{N}$ such that given any $(\what{w}, w, z) \in W \times W \times Z$, where $\what{w} \neq w$, the processes $\{\what{w}_k\}$ and $\{(w_k,z_k)\}$ can follow, resp., $\kappa$-length paths $\what{w} = \what{h}^0,\dots, \what{h}^{\kappa} = w'$, and 
        $(w,z)= (h^0,g^0), \dots, (h^{\kappa},g^{\kappa}) = (w',z')$ for some $z'$, where $\what{h}^l \neq h^l$ for all $l <\kappa$. Furthermore, for all $k$, we have
        \be\label{eq:cond1bound}
            \Pr\left( \bigwedge_{l=1}^{\kappa} \what{w}_{k+l} = \what{h}^{l} \mid \what{w}_k = \what{w} \right) \geq \hat{\varepsilon},\Pr\left( \bigwedge_{l=1}^{\kappa} w_{k+l} = h^{l} \mid (w_{k}, z_{k}) = (w, z) \right) \geq \varepsilon.
        \ee
    
    \textbf{Condition $(ii)$} For any $k=0,\ldots,K-1$, there exists $\lambda \in (0,1]$ such that
        \be
        \TV{\mathbb{P}^w \left( \cdot \mid (w_l, z_l)_{l \leq k} \right) - \what{\mathbb{P}}(\cdot \mid w_k)} \leq 1-\lambda \quad \forall (w_l, z_l)_{l \leq k}.
        \ee

        Let $\what{\mu}_k \in \Delta(W)$ and $\mu_k \in \Delta(W \times \mathbb{Z})$ denote, resp., the distributions of $\what{w}_k$ and $(w_k, z_k)$. Furthermore, $\what{\mu}^w \in \Delta(W)$ denotes the marginal of $\mu_k$ over $Z$. Then, there exists a coupling over $\{\what{w}_k\}$ and $\{(w_k, z_k)\}$ such that
    \color{black}
        \be
            \|\what{\mu}_k - \mu_k^w\|_1 \leq 2(1-\varepsilon\what{\varepsilon})^{\frac{k}{\kappa}-1} + 2(1-\lambda^{\kappa})\frac{1+\varepsilon\what{\varepsilon}}{\varepsilon\what{\varepsilon}}\quad \forall t =0,\ldots,K-1.\label{eq:result1}
        \ee
    \end{lemma}
    
    We highlight that the first term on the right-hand side of \eqref{eq:result1} goes to zero as $k\rightarrow\infty$ whereas the second term is a constant depending on the distance between the transition probabilities. For example, if $\lambda = 1$ for $K\rightarrow \infty$, then the processes $\{\what{w}_k\}$ and $\{w_k\}$ are identical Markov chains and the second term is zero. Correspondingly, the proof of \Cref{lem:coupling} (\Cref{sec:coupling}) has a flavor similar to the proof of the Ergodicity Theorem for Markov chains \citep[Chapter 4]{ref:Levin17}. 
    
    \begin{remark}
    Existing perturbation bounds for the stationary distribution of Markov chains focus mainly on homogeneous chains using maximum entry-wise perturbations \citep{ref:Cho01,ref:Liu12}. However, these bounds are not tight enough for our analysis. In \Cref{lem:coupling}, we exploit the structure of these chains to derive sharper bounds and address non-homogeneous cases through finite-time analysis, making \Cref{lem:coupling} independently interesting.
    \end{remark}
    
    Lastly, the following proposition characterizes the stationary distributions of the action profiles in the classical and independent learning dynamics based on \Cref{lem:coupling}. A preliminary version of this lemma appeared in \citep{ref:Onur23}.
    
    \begin{proposition}\label{lem:stationary}
    Consider that agents follow the logit dynamics in the repeated play of a potential game $\mathcal{G}=\langle (A^i,u^i)_{i\in [N]}\rangle$ with the potential function $\Phi$. Let $\mu^{\mathrm{coor}},\mu^{\mathrm{ind}}\in \Delta(A)$ denote the unique stationary distributions of the Markov chains formed by the actions profiles for the coordinated and independent revision processes, as described in \eqref{eq:Pcoor}. Then, we have $\mu^{\mathrm{coor}} = \br(\Phi(\cdot))$ and
    \be\label{eq:resultstationary}
    \|\mu^{\mathrm{coor}}-\mu^{\mathrm{ind}}\|_1 \leq \Lambda(\delta) := 2(1-\lambda(\delta))\cdot \frac{1+\lambda(\delta)\cdot\epsilon^2}{\lambda(\delta)\cdot \epsilon^2}.
    \ee
    For independent-revision schemes, $\lambda(\delta) \in (0,1)$ is the probability that only one agent revises actions for $N$ stages conditioned on that at least one agent revises actions and $\epsilon\in(0,1)$ is a uniform lower bound on the probability for any $N$-length trajectory where only one agent revises actions, which depends on the inherent exploration.  
    \end{proposition}
    
    As $\delta\rightarrow 0^+$, we have $\Lambda(\delta)\rightarrow 0$ since $\lambda(\delta)\rightarrow 1$. Note the resemblance of \eqref{eq:resultstationary} to the second term on the right-hand side of \eqref{eq:result1}. Correspondingly, the proof of \Cref{lem:stationary} (\Cref{sec:stationarydist}) follows from \Cref{lem:coupling} by quantifying the distance between the transition matrices of the coordinated and independent-revision schemes.
    
    In the following, we leverage \Cref{prop:propConstantPayoff,lem:stationary} and \Cref{lem:coupling} to prove that the estimated plays can track the (soft) maximizers of the Q-function estimates.
    
    \subsection{Tracking Result for the Estimated Plays} \label{sec:division}
    For the estimated plays, we can focus on $\piave_t$ since we can compute $\pifree_t$ based on $\piave_t$ according to
    \be\label{eq:pipi}
    \pifree_t(s)(a) > 0 \quad\Leftrightarrow \quad a\in A_t(s) = \argmax_{\tilde{a}\in A} \{\piave_t(s)(a)\}\quad\forall (s,a).
    \ee
    Let $\mu(u)\in\Delta(A)$ denote the unique stationary distribution of the action profiles for the logit dynamics followed in the identical-interest game $\mathcal{G}=\langle (A^i)_{i\in [N]},u\rangle$. Then, the following proposition shows that the empirical averages $\piave_t$ can track the stationary distributions associated with the stage-game payoffs $Q_t$'s. 
    
    \begin{proposition}[Tracking Result]\label{prop:tracking}
    We have $\|\piave_t(s)-\mu(Q_{t}(s,\cdot))\|_1 \rightarrow 0$ as $t\rightarrow\infty$ almost surely for each $s$.
    \end{proposition}
    
    \begin{proof}

    If the stage-game payoffs, i.e., the Q-function estimates,  remained fixed over time, the empirical distribution of action profiles, \(\piave_t\), would converge to the stationary distribution of the homogeneous Markov chain governing these profiles, thanks to the ergodicity of the chain.
    
    In our setting, however, the Q-function estimates evolve as the game is played. On the other hand, the use of a vanishing step size (as specified in \Cref{assume:stepsize}-\ref{subasssume:i}) and the boundedness of the iterates by the nature of the update \eqref{eq:Q} yield that \textit{one-step} update to the Q-function becomes progressively smaller over time, decaying to zero.
    
    Because the logit dynamics we employ are designed such that small changes in the payoffs (Q-functions) induce also (relatively) small changes in the transition probabilities (as established in Proposition \(\ref{prop:propConstantPayoff}\)), the stage games behave almost as if they are stationary over short time intervals. In these ``quasi-stationary" periods, the changes accumulated in the transition probabilities are negligible enough that the system's short-run behavior mimics that of a stationary process.
    
    Over longer time periods, even these small changes can add up, leading to significant shifts in the dynamics. To address this, we partition the overall time horizon into epochs whose lengths grow at a sufficiently slow rate. More explicitly, fix some state $s$ and divide the horizon into epochs such that the state $s$ gets visited only for a certain number of times specific to the epoch and the first state of every epoch is $s$. For clear exposition, we use $k$ and $(n)$, resp., as subscripts for the parameters associated with the $k$th visit to $s$ and the $n$th epoch for $s$. For example, $t_k$, $t_{(n)}$, and $k_{(n)}$ denote the stage that $s$ gets visited for the $k$th time, the stage that the $n$th epoch starts, and the number of times $s$ gets visited until and including stage $t_{(n)}$, respectively.
    \color{black}
    
    For each epoch $n$, we define the filtration $\mathcal{F}_{(n)} = \sigma(\{s_t,a_t\}_{t<t_{(n)}},s_{t_{(n)}})$ and 
    \be\label{eq:muk}
    \mu_k := \mathbb{E}[a_{t_k}\mid \mathcal{F}_{(n)}]\in \Delta(A)
    \ee 
    denotes the distribution over the action profiles played at the $k$th visit to $s$ conditioned on $\mathcal{F}_{(n)}$. We can invoke \Cref{lem:coupling} to characterize the distribution $\mu_k$ for $k=k_{(n)},\ldots,k_{(n+1)}-1$ within epoch $n$ in comparison to the stationary distribution of the logit dynamics in a \textit{fictional} (stationary) scenario. To this end, we focus on the evolution of the action profiles across consecutive visits to state $s$ while modeling the state-action trajectories in between the consecutive visits to $s$ as some external (possibly coupled) processes, as illustrated in \Cref{fig:processes}. 
    
    Let $Q_{(n)}(\cdot):=Q_{t_{(n)}}(\cdot)$ and $c_{(n)}(\cdot):=c_{t_{(n)}}(\cdot)$.\footnote{We drop the agent index $i$ in $Q_t^i$ since the agents have common estimates.} For the fictional scenario, we consider the logit dynamics followed in the repeated play of $\mathcal{G}_{(n)}=\langle (A^i)_{i\in [N]}, Q_{(n)}(s,\cdot)\rangle$ at every visit to state $s$ in the main scenario. Correspondingly, $\mu(Q_{(n)}(s,\cdot))\in\Delta(A)$ is the unique stationary distribution of the action profiles for the logit dynamics followed in $\mathcal{G}_{(n)}$. On the other hand, by \eqref{eq:Q} and \Cref{assume:stepsize}-\ref{subasssume:i}, we can bound the changes on the Q-function estimates within epoch $n$ in the main scenario by
    \be\label{eq:QQbound}
    |Q_{t}(s,a) - Q_{(n)}(s,a)|\leq H K_{(n)}\beta_{(n)} \quad\forall a\in A,\; t\in [t_{(n)},t_{(n+1)}),
    \ee
    
    where we define $\beta_{(n)}\coloneqq \max_{a}\{\beta_{c_{(n)}(s,a)}\}$, $K_{(n)}\coloneqq k_{(n+1)}-k_{(n)}$, and
    $H \coloneqq 2/(1-\gamma)$ is a uniform upper bound on $|r_{t}-\gamma v_t(s_{t+1})-Q_t(s,a)|$ for all $(s,a)$ and $t$.
    \color{black}
    
    The following lemma characterizes $\mu_k$ in terms of $\mu(Q_{(n)}(s,\cdot))$ by coupling the evolution of the dynamics in the main and fictional scenarios based on \eqref{eq:QQbound}.
    
    \begin{lemma}\label{lem:lambda}
    For sufficiently large $n$, we have 
    \be\label{eq:lambdaresult}
    \|\mu_{k} - \mu(Q_{(n)}(s,\cdot))\|_1 \leq C \rho^{k-k_{(n)}} + D K_{(n)} \beta_{(n)}\quad\forall k\in[k_{(n)},k_{(n+1)})
    \ee
    for some constants $C,D>0$ and $\rho\in (0,1)$.
    \end{lemma}

    The proof (\Cref{sec:lambda}) follows from \Cref{prop:propConstantPayoff} and \Cref{lem:coupling}. We highlight the resemblance of \eqref{eq:lambdaresult} to \eqref{eq:result1}.
    \color{black}
    
    By \eqref{eq:piave}, the empirical average $\piave_t(s)$ evolves according to
    \be\label{eq:PiUpdate}
        \piave_{t+1}(s) = \piave_{t}(s) + \mathds{1}_{\{s_t=s\}} \cdot \frac{1}{c_t(s)} \cdot (a_t - \piave_{t}(s)).
    \ee
    Henceforth, we view actions $a\subset \Delta(A)$ as pure strategies, or probability vectors, in which the associated action gets played with probability $1$. By \eqref{eq:PiUpdate}, $\piave_t(s)$ gets updated only when $s$ gets visited, i.e., $\piave_t(s) = \piave_{t_k}(s)$ for all $t\in [t_k,t_{k+1})$. Then, we can write \eqref{eq:PiUpdate} as
    \be\label{eq:pik}
    \piave_{t_{k+1}}(s) = \piave_{t_k}(s) + \alpha_k \cdot (a_{t_k} - \piave_{t_k}(s)),
    \ee
    where $\alpha_k := 1/k$. 
    
    The following lemma shows that accumulated changes can still form a stochastic approximation recursion for certain (growing) epoch lengths.
    
    \begin{lemma}\label{lem:stepsizein}
        Consider the iterative update rule \eqref{eq:pik} with some step sizes $\{\alpha_k\in [0,1]\}$ satisfying the standard conditions that $\alpha_k\rightarrow 0$ monotonically, $\sum_k \alpha_k = \infty$ and $\sum_k \alpha_k^2 < \infty$. Then, \eqref{eq:pik} yields that we can write the accumulated changes in $\piave_{(n)}(s) := \piave_{t_{(n)}}(s)$ across epochs $n=1,2,\ldots$ as
    \be
    \piave_{(n+1)}(s) = \piave_{(n)}(s) + \oalpha_{(n)}  \cdot \left( \sum_{k=k_{(n)}}^{k_{(n+1)}-1} \frac{\oalpha_k}{\oalpha_{(n)}} \cdot a_{t_k} - \piave_{(n)}(s) \right), \label{eq:epochupdatePi}
    \ee
    where the auxiliary step sizes are defined by
    \be\label{eq:alphaLemma}
    \oalpha_{(n)} := \sum_{k=k_{(n)}}^{k_{(n+1)}-1} \oalpha_k\quad\mbox{and}\quad\oalpha_k:= \alpha_k \prod_{l=k+1}^{k_{(n+1)}-1}(1-\alpha_l)\quad\forall k\in [k_{(n)},k_{(n+1)}).
    \ee
    
    Furthermore, there exist epoch lengths $K_{(n)}:=k_{(n+1)}-k_{(n)}$ growing at a sufficiently slow rate that $K_{(n)}\rightarrow \infty$ and $K_{(n)}/k_{(n)} \rightarrow 0$ as $n\rightarrow\infty$ while the step sizes $\{\oalpha_{(n)} \in [0,1]\}$ satisfy the standard conditions that $\oalpha_{(n)}\rightarrow 0$, $\sum_n \oalpha_{(n)}=\infty$, and $\sum_n \oalpha_{(n)}^2 <\infty$.
    \color{black}
    \end{lemma}

    As an illustrative example for \Cref{lem:stepsizein}, consider $L_n \coloneqq \lfloor \log_2(n)\rfloor$ and a step size sequence given by $\alpha_k = 1/k$. Define the epoch lengths as $K_{(n)} = 1+L_n$. Then, we have 
    $k_{(n+1)} = nL_n + L_n + n - 2^{L_n + 1}+2 \geq n (L_n -1)$,
    which leads to the accumulated step size estimate $\oalpha_{(n)}=K_{(n)}/(k_{(n+1)}-1)$. From this, we derive the bounds  $L_n/(n (L_n +1)) \leq \oalpha_{(n)} \leq L_n/(n (L_n -1))$. Taking the limit as $n\rightarrow\infty$, we obtain $\lim_{n \rightarrow \infty}K_{(n)}/k_{(n)} = 0$, and $\lim_{n \rightarrow \infty}\oalpha_{(n)} = 0$. Moreover, the accumulated step size $\oalpha_{(n)}$ behaves asymptotically like $1/n$, ensuring that $\sum_n \oalpha_{(n)}=\infty$, and $\sum_n \oalpha_{(n)}^2 <\infty$. \Cref{lem:stepsizein} is a general result asserting that for any step size sequence $\alpha_k$ satisfying the lemma's conditions, we can always determine an appropriate epoch length $K_{(n)}$. The proof of \Cref{lem:stepsizein} is provided in \Cref{sec:stepsizein}.
    \color{black}
    
    For $\alpha_k=1/k$, the accumulated update \eqref{eq:epochupdatePi} can be written as
    \be\label{eq:Piaccumulated}
    \piave_{(n+1)}(s) = \piave_{(n)}(s) + \oalpha_{(n)}  \cdot \left( \mu(Q_{(n)}(s,\cdot)) + e_{(n+1)} + \omega_{(n+1)} - \piave_{(n)}(s)\right),
    \ee
    where $\oalpha_{(n)}= K_{(n)}/(k_{(n+1)}-1)$ by \eqref{eq:alphaLemma} and the auxiliary error terms are defined by
    \be
    e_{(n+1)} := \frac{1}{K_{(n)}}\sum_{k=k_{(n)}}^{k_{(n+1)}-1} \hspace{-.1in}(\mu_k - \mu(Q_{(n)}(s,\cdot))),\quad\omega_{(n+1)}:= \frac{1}{K_{(n)}}\sum_{k=k_{(n)}}^{k_{(n+1)}-1} \hspace{-.1in}(a_{t_k}-\mu_k).
    \ee
    Both $e_{(n+1)}$ and $\omega_{(n+1)}$ depend on $s$ implicitly for notational simplicity. The bound on $\|\mu_k-\mu(Q_{(n)}(s,\cdot))\|$ in \Cref{lem:lambda} yields that the error $e_{(n+1)}$ is bounded by
    \be\label{eq:errorboundbeta}
    \left\|e_{(n+1)}\right\|_1\leq \frac{C}{1-\rho}\cdot \frac{1}{K_{(n)}} + D\cdot K_{(n)}\beta_{(n)}
    \ee
    for sufficiently large $n$, due to the triangle inequality. On the other hand, $\omega_{(n+1)}$ forms a square-integrable Martingale difference sequence by the definition \eqref{eq:muk}.
    
    At the same timescale with \eqref{eq:Piaccumulated}, we can write the accumulated changes for the Q-update \eqref{eq:Q} as
    \be\label{eq:Qaccumulated}
    Q_{(n+1)}(s, \cdot) = Q_{(n)}(s, \cdot) + \oalpha_{(n)}  \cdot  \epsilon_{(n+1)} 
    \ee
    for some error term $\epsilon_{(n+1)}\in\mathbb{R}^{|A|}$, bounded by 
    \be\label{eq:epsilonboundbeta}
    \|\epsilon_{(n+1)}\|_\infty \leq H\beta_{(n)} k_{(n+1)} 
    \ee 
    due to \eqref{eq:QQbound}.
    
    The following lemma shows that the error terms are asymptotically negligible.
    
    \begin{lemma}\label{lem:ane}
    We have $\beta_{(n)}k_{(n+1)}\rightarrow 0$, and therefore, $\|e_{(n)}\|\rightarrow 0$ and $\|\epsilon_{(n)}\|\rightarrow 0$ as $n\rightarrow\infty$ almost surely.
    \end{lemma}
    
    \begin{proof}
    Recall that each agent revising their action chooses any available action with a positive probability that is uniformly bounded away from zero, and every agent revises their action with some fixed positive probability. Since there are finitely many states, at least one state is visited infinitely often; call this state $s$. Because logit dynamics allow one action change per agent at a time (as is also possible in the independent case), any action profile can be realized with positive probability after $s$ is visited $N$ times. Therefore, for any 
    $a \in A$, the state-action pair $(s,a)$ is realized infinitely often, and the frequencies of played actions are comparable. Moreover, by the irreducibility assumption stated in \Cref{thm:main}, there is at least one action profile $a$ that leads from $s$ to some reachable state $s'$. Since $(s,a)$ is visited infinitely often, it follows that $s'$ is also visited infinitely often. By induction, and given that agents can play any action profile at comparable frequencies, the irreducibility assumption in the statement of \Cref{thm:main} implies that every state $s$ is visited infinitely often i.e., $c_t(s)\rightarrow\infty$ as $t\rightarrow\infty$ almost surely. Correspondingly, any pair $(s,a)$ gets realized infinitely often, i.e., $c_t(s,a)\rightarrow\infty$ as $t\rightarrow\infty$ almost surely. Therefore, if we follow similar lines with the proof of \citep[Lemma 6.1]{ref:Sayin20}, \Cref{assume:stepsize}-\ref{subasssume:iii} yields that 
    \be\label{eq:Borel}
    \lim_{t\rightarrow\infty}c_t(s) \beta_{c_{t}(s,a)} = 0 \quad\forall (s,a),
    \ee  
    with probability $1$.\color{black} 
    
    Recall also that $k_{(n)}=c_{(n)}(s)$ and $c_{(n)}(s,a)$ denote, resp., the number of times $s$ and $(s,a)$ get realized in the first stage of epoch $n$. By \eqref{eq:Borel}, $k_{(n)}\beta_{c_{(n)}(s,a)}$ decays to zero for all $(s,a)$, and therefore, $k_{(n)}\beta_{(n)}$ decays to zero since there are finitely many pairs and $\beta_{(n)}=\max_{a}\{\beta_{c_{(n)}(s,a)}\}$. 
    
    Let $K_{(n)}$ be chosen according to \Cref{lem:stepsizein}. Then, such a choice of $K_{(n)}$ satisfies $k_{(n)} \geq n \geq K_{(n)}$ as $\{K_{(n)}\}$ grows at a sub-linear rate and $K_{(n)} \geq 1$ for all $n$. Therefore, $K_{(n)}\beta_{(n)}$ and $k_{(n+1)}\beta_{(n)}=K_{(n)}\beta_{(n)}+k_{(n)}\beta_{(n)}$ decay to zero almost surely. By \eqref{eq:errorboundbeta} and \eqref{eq:epsilonboundbeta}, the error terms also decay to zero almost surely.
    \end{proof}
    
    \Cref{lem:ane} yields that we can rewrite the accumulated changes \eqref{eq:Piaccumulated} and \eqref{eq:Qaccumulated}, resp., on $\piave_{(n)}(s)$ and $Q_{(n)}(s,\cdot)$ across epochs as
    \begin{subequations}\label{eq:stochasticApprDiscrete}
    \begin{flalign} 
    &\piave_{(n+1)}(s) = \piave_{(n)}(s) + \oalpha_{(n)}\cdot(\mu(Q_{(n)}(s,\cdot)) - \piave_{(n)}(s) + \omega_{(n+1)} + \ane)\\
    &Q_{(n+1)}(s,\cdot) = Q_{(n)}(s,\cdot) + \oalpha_{(n)}\cdot \ane,
    \end{flalign}
    \end{subequations}
    where $\ane$ refers to asymptotically negligible error terms. Since step sizes $\oalpha_{(n)}$ satisfies the stochastic approximation conditions by \Cref{lem:stepsizein}, the function $\mu(\cdot)$ is locally Lipschitz continuous by \Cref{prop:propConstantPayoff}, $\omega_{(n+1)}$ forms a square-integrable Martingale difference sequence, and the iterates $(\piave_{(n)}(s),Q_{(n)}(s,\cdot))$ are bounded, we can approximate the discrete-time accumulated updates \eqref{eq:stochasticApprDiscrete} with the following continuous-time flow
    \be\label{eq:ode}
        \dot{\pi}= \mu(Q) -\pi \quad\mbox{and}\quad\dot{Q} = 0
    \ee
    such that \eqref{eq:stochasticApprDiscrete} converges to an internally chain-recurrent set of \eqref{eq:ode} \citep{ref:Benaim99}. Furthermore, we can characterize its internally chain-recurrent set by formulating a Lyapunov function \citep{ref:Benaim99}. For example, the Lyapunov function $V(\pi,Q) = \|\mu(Q) -\pi\|_2^2$ yields that
    \begin{flalign}\label{eq:limitPi}
        \lim_{n \rightarrow 0} \|\piave_{(n)}(s)-\mu(Q_{(n)}(s,\cdot))\|_1  = 0
    \end{flalign}
    almost surely as norms are comparable in finite-dimensional spaces.
    
    We have the tracking result \eqref{eq:limitPi} for the first stages $t_{(n)}$ of each epoch $n$. We can also show the tracking result for each $t$ since  
    \begin{flalign}
        \|\piave_t(s) - \mu(Q_t(s,\cdot))\|_1 \leq &\;\|\piave_t(s) - \piave_{(n)}(s)\|_1 + \| \mu(Q_{(n)}(s,\cdot)) -  \mu(Q_{t}(s,\cdot))\|_1\nn\\
        &+ \|\piave_{(n)}(s) - \mu(Q_{(n)}(s,\cdot))\|_1\nn
    \end{flalign}
    for all $t \in [t_{(n)}, t_{(n+1)})$, by the triangle inequality. The first term on the right-hand side is bounded by
    \begin{flalign} 
        \|\pi_t(s) - \pi_{(n)}(s)\|_1 \leq 2 K_{(n)}/{k_{(n)}}\quad\forall t \in [t_{(n)}, t_{(n+1)}) \label{eq:pi_t_nbound}
    \end{flalign}
    by \eqref{eq:PiUpdate} and decays to zero by \Cref{lem:stepsizein}. The second one is bounded by 
    \begin{flalign} \label{eq:mu_t_mu_nbound}
        \|\mu(Q_t(s,\cdot)) - \mu(Q_{(n)}(s,\cdot))\|_1 \leq LHK_{n}\beta_{(n)}\quad\forall t \in [t_{(n)}, t_{(n+1)})
    \end{flalign}
    for some constant $L$, by \Cref{prop:propConstantPayoff} and \eqref{eq:QQbound}, and decays to zero by \Cref{lem:ane}. The last one decays to zero by \eqref{eq:limitPi}. This completes the proof of \Cref{prop:tracking} since the fixed state $s$ is arbitrary.
    \end{proof}
    
    \subsection{Convergence of the Q-function Estimates}
    
    The convergence of the Q-function estimates to $Q_*$, as described in \eqref{eq:Qglobal}, follows from showing that the value function estimates track the joint maximizer of the Q-function estimates approximately, as shown in the following corollary to \Cref{prop:tracking}.  
    
    \begin{corollary}\label{cor:tracking}
    We have $\limsup_{t\rightarrow\infty} |v_t(s) - \max_a \{Q_t(s,a)\}| \leq e_Q\cdot(1-\gamma)/\gamma$ almost surely, where $e_Q\geq 0$ is as described in \Cref{tab:bounds}.
    \end{corollary}
    
    \begin{proof}
    For AL-Q, we have $\pi_t \equiv \piave_t$ and 
    $\|\pi_t(s)-\mu^{\mathrm{coor}}(Q_t(s,\cdot))\|_1 = \|\pi_t(s)-\br(Q_t(s,\cdot))\|_1\rightarrow 0$ as $t\rightarrow\infty$ by \Cref{lem:stationary,prop:tracking}. This implies that 
    \be
    \limsup_{t\rightarrow\infty} |v_t(s) - \max_a \{Q_t(s,a)\}|\leq \tau\log|A|
    \ee
    since $|\max_a\{Q_t(s,a)\} - \mathbb{E}_{a\sim\br(Q_t(s,\cdot))}[Q_t(s,a)]\,| \leq \tau\log|A|$.
    
    For AIL-Q, we have $\pi_t \equiv \piave_t$ and 
    $\|\pi_t(s)-\mu^{\mathrm{ind}}(Q_t(s,\cdot))\|_1\rightarrow 0$ as $t\rightarrow\infty$ by \Cref{prop:tracking}. By the triangle inequality, we have
    \begin{flalign}
    |v_t(s) - \max_a \{Q_t(s,a)\}|\leq &\;|\mathbb{E}_{a\sim\pi_t(s)}[Q_t(s,a)] - \mathbb{E}_{\mu^{\mathrm{ind}}(s)}[Q_t(s,a)]|\\
    &|\mathbb{E}_{\mu^{\mathrm{ind}}(s)}[Q_t(s,a)] - \mathbb{E}_{\mu^{\mathrm{coor}(s)}}[Q_t(s,a)]|\nn\\
    &|\mathbb{E}_{\mu^{\mathrm{coor}(s)}}[Q_t(s,a)] - \max_a \{Q_t(s,a)\}|.
    \end{flalign}
    The first term on the right-hand side decays to zero while the last is bounded by $\tau\log|A|$. If we invoke \Cref{lem:stationary} for the second term, we obtain
    \be
    \limsup_{t\rightarrow\infty} |v_t(s) - \max_a \{Q_t(s,a)\}|\leq \tau\log|A| + \frac{\Lambda(\delta)}{1-\gamma}
    \ee
    since the Q-function estimates are bounded by $1/(1-\gamma)$.
    
    For EIL-Q, we have $\pi_t \equiv \pifree_t$ and 
    \begin{flalign}
    \|\piave_t(s)-\br(Q_t(s,\cdot))\| \leq &\;\|\piave_t(s) - \mu^{\mathrm{ind}}(Q_t(s,\cdot))\| \\
    &+ \|\mu^{\mathrm{ind}}(Q_t(s,\cdot)) - \mu^{\mathrm{coor}}(Q_t(s,\cdot))\|\nn
    \end{flalign}
    for any norm by the triangle inequality. Then, \Cref{prop:tracking,lem:stationary} yield
    \be\label{eq:pi_br_dist}
        \limsup_{t \rightarrow \infty} \| \piave_t(s) - \br(Q_t(s,\cdot))\|_{\infty} \leq \Lambda(\delta).
    \ee
    as $\|\cdot\|_{\infty}\leq \|\cdot\|_1$. Let $a^1\in\argmax_a \{\br(Q_t(s,\cdot))(a)\}$ and $a^2\in\argmax_a\{\piave_t(s)(a)\}$. Then, we can bound the difference of the probabilities of these action profiles in $\br(Q_t(s,\cdot))$ by
    \begin{flalign} \label{eq:qa1a2bound}
        0 & \leq  \br(Q_t(s,\cdot))(a^1) - \br(Q_t(s,\cdot))(a^2) \\
        & \leq \br(Q_t(s,\cdot))(a^1) - \piave_t(s)(a^1) + \piave_t(s)(a^2) - \br(Q_t(s,\cdot))(a^2)\nn\\
        & \leq 2\Lambda(\delta) + \ane,\nn
    \end{flalign}
    where the first and second inequalities follow since $a^1$ and $a^2$ are, resp., the maximizers of the distributions $\br(Q_t(s,\cdot))$ and $\piave_t(s)$, and the last inequality follows from \eqref{eq:pi_br_dist}.
    
    The boundedness of the Q-iterates by $1/(1-\gamma)$ and the smoothed best response \eqref{eq:logit} guarantee that 
    \be\label{eq:minprob}
    \br(Q_t(s,\cdot))(a) \geq \epsilon := \frac{1}{|A|}\exp\left(-\frac{1}{\tau(1-\gamma)}\right)\quad\forall (s,a).
    \ee
    Therefore, we can bound the difference of these action profiles in $Q_t(s,\cdot)$ by
    \begin{flalign}
    Q_t(s,a^1)-Q_t(s,a^2) &= \tau \cdot \log\left(\frac{\br(Q_t(s,\cdot))(a^1)}{\br(Q_t(s,\cdot))(a^2)}\right)\\
    &\leq \tau \cdot \log\left(1+2\Lambda(\delta)/\epsilon + \ane\right),\label{eq:Lambdaane}
    \end{flalign}
    where the first line follows from \eqref{eq:logit}, and the second from \eqref{eq:qa1a2bound} and \eqref{eq:minprob}. Since $\log(\cdot)$ is monotonically increasing and continuous, \eqref{eq:Lambdaane} yields that
    \begin{flalign} \label{eq:defXi}
        \limsup_{t\rightarrow \infty} \{Q_t(s,a^1)-Q_t(s,a^2)\} \leq \Xi(\delta) \coloneqq \tau \cdot\log \left(1 + 2\Lambda(\delta)/\epsilon\right).
    \end{flalign}
    The inequality \eqref{eq:defXi} holds for any $a^2 \in A_t(s)$ since $A_t(s)$, as described in \eqref{eq:pipi}, is the set of maximizers of $\piave_t(s)$. Moreover, by the definition \eqref{eq:logit}, $Q_t(s,a^1) = \max_a \{Q_t(s,a)\}$. Since any convex combination of $Q_t(s,a^2)$ over $a^2 \in A_t$, including $\pifree_t(s)\in\Delta(A)$, also satisfies the inequality \eqref{eq:defXi}, we obtain
    \be\label{eq:EILbound}
        \limsup_{t\rightarrow \infty} |v_t(s) - \max_a \{Q_t(s,a)\}| \leq \Xi(\delta)
    \ee
    almost surely, as $v_t(s) = \mathbb{E}_{a\sim\pifree_t(s)}[Q_t(s,a)]$.
    
    For EL-Q, the right-hand side of \eqref{eq:pi_br_dist} is $0$ by \Cref{prop:tracking}. Hence, \eqref{eq:EILbound} yields that $|v_t(s) - \max_a \{Q_t(s,a)\} - v_t(s)|$ decay to zero almost surely.
    \end{proof}
    
    We can write the Q-update \eqref{eq:Q} as
    \begin{flalign}
    Q_{t+1}(s,a) = &\;Q_t(s,a) + \mathds{1}_{\{(s,a)=(s_t,a_t)\}}\beta_{c_t(s,a)} \\
    &\times\Big(r(s,a) + \gamma \sum_{s'}p(s'\mid s,a)\max_{\tilde{a}} \{Q_t(s',\tilde{a})\} - Q_t(s,a) \nn\\
    &+ \bar{\epsilon}_t(s,a) + \bar{\omega}_{t+1}(s,a)\Big),\nn
    \end{flalign}
    where the error term and the stochastic approximation noise are, resp., defined by
    \begin{subequations}
    \begin{flalign}
    &\bar{\epsilon}_t(s,a) = \gamma \sum_{s'}p(s'\mid s,a)\big(v_t(s') - \max_{\tilde{a}} \{Q_t(s',\tilde{a})\}\big)\label{eq:barepsilon}\\
    &\bar{\omega}_{t+1}(s,a) = \gamma\left(v_t(s_{t+1}) - \sum_{s'}p(s'\mid s,a)v_t(s')\right).
    \end{flalign}
    \end{subequations}
    \Cref{cor:tracking} and \eqref{eq:barepsilon} yield that the error term $\bar{\epsilon}_t$ satisfies
    \be
    \limsup_{t\rightarrow\infty}|\bar{\epsilon}_t(s,a)|\leq (1-\gamma)e_Q\quad\forall (s,a)
    \ee
    almost surely. On the other hand, since $s_{t+1}\sim p(\cdot\mid s_t,a_t)$, the stochastic approximation noise $\bar{\omega}_{t+1}$ forms a square-integrable Martingale difference sequence. Due to the exploration inherent to the logit-Q dynamics, the irreducibility assumption in the statement of \Cref{thm:main} yields that each $(s,a)$ pair gets realized infinitely often. Based on the contraction property of the Bellman optimality operator, Theorem 5.1 in the extended version of \citep{ref:Sayin20} yields \eqref{eq:result}.\footnote{The extended version of \citep{ref:Sayin20} is available at \url{https://arxiv.org/abs/2010.04223}.}
    
    \subsection{Convergence of the Estimated Plays to Equilibrium}
    
    Recall that $v_{\pi}:S\rightarrow\mathbb{R}$ and $Q_{\pi}:S\times A\rightarrow\mathbb{R}$, as described in \eqref{eq:local}, resp., denote the value and Q-function of the Markov stationary policy $\pi:S\rightarrow\Delta(A)$. For example, we have $v_{\pi}(s) = \E_{a\sim\pi(s)}[Q_{\pi}(s,a)]$. Based on $Q_*$, as described in \eqref{eq:Qglobal}, we define $v_*(s) = \max_a\{Q_*(s,a)\}$ for all $s$. Then, given the strategy $\pi_t$, we have
    \begin{align}
    0 &\leq v_*(s) - v_{\pi_t}(s)\\
    &=  v_*(s) -  \E_{a\sim\pi_t(s)}[Q_*(s,a)] + \E_{a\sim\pi_t(s)}[Q_*(s,a) - Q_{\pi_t}(s,a)].\label{eq:em2}
    \end{align}
    Since $(v_*,Q_*)$ and $(v_{\pi_t},Q_{\pi_t})$ pairs satisfy the one-step Bellman equation, we also have
    \begin{align}
    \E_{a\sim\pi_t(s)}[Q_*(s,a) - Q_{\pi_t}(s,a)] &= \gamma\E_{a\sim\pi_t(s)}\left[\sum_{s'\in S}p(s'\mid s,a) (v_{*}(s') - v_{\pi_t}(s'))\right]\\
    &\leq \gamma \|v_*-v_{\pi_t}\|_{\infty}\label{eq:em3}
    \end{align}
    for both $\pi_t \equiv \piave_t$ and $\pi_t \equiv \pifree_t$. On the other hand, we can bound the first two terms $v_*(s) -  \E_{a\sim\pi_t(s)}[Q_*(s,a)]$ on the right-hand side of \eqref{eq:em2} by
    \begin{flalign}
        |v_*(s)-&\E_{a\sim \pi_{t}(s)}[Q_*(s,a)]| \leq |\max_{a}\{Q_*(s,a)\}-\max_{a}\{Q_t(s,a)\}|\\
        &+|\max_{a}\{Q_t(s,a)\}-v_t(s)|+|\E_{a\sim \pi_{t}(s)}[Q_*(s,a)] - \E_{a\sim \pi_{t}(s)}[Q_t(s,a)]| \nn
    \end{flalign}
    Then, we can bound the first and third terms on the right-hand side by \eqref{eq:result} and bound the second term by \Cref{cor:tracking}, and obtain
    \be\label{eq:diffboundeQ}
    \limsup_{t\rightarrow\infty}|v_*(s)-\E_{a\sim \pi_{t}(s)}[Q_*(s,a)]|\leq \frac{1+\gamma}{\gamma}\cdot e_Q
    \ee
    due to the sub-additivity of the limit superior. Since the upper bound in \eqref{eq:diffboundeQ} does not depend on $(s,a)$, \eqref{eq:em2}, \eqref{eq:em3}, and \eqref{eq:diffboundeQ} yields that
    \be
    \limsup_{t\rightarrow\infty}\|v_*-v_{\pi_t}\|_{\infty} \leq \frac{1+\gamma}{\gamma(1-\gamma)} e_Q.
    \ee
    This completes the proof of \Cref{thm:main}.

    \begin{figure}[t!] 
        \centering

        \begin{subfigure}[b]{0.525\textwidth}
    \color{black}
            \centering
    
            \includegraphics[width=\textwidth,page=4]{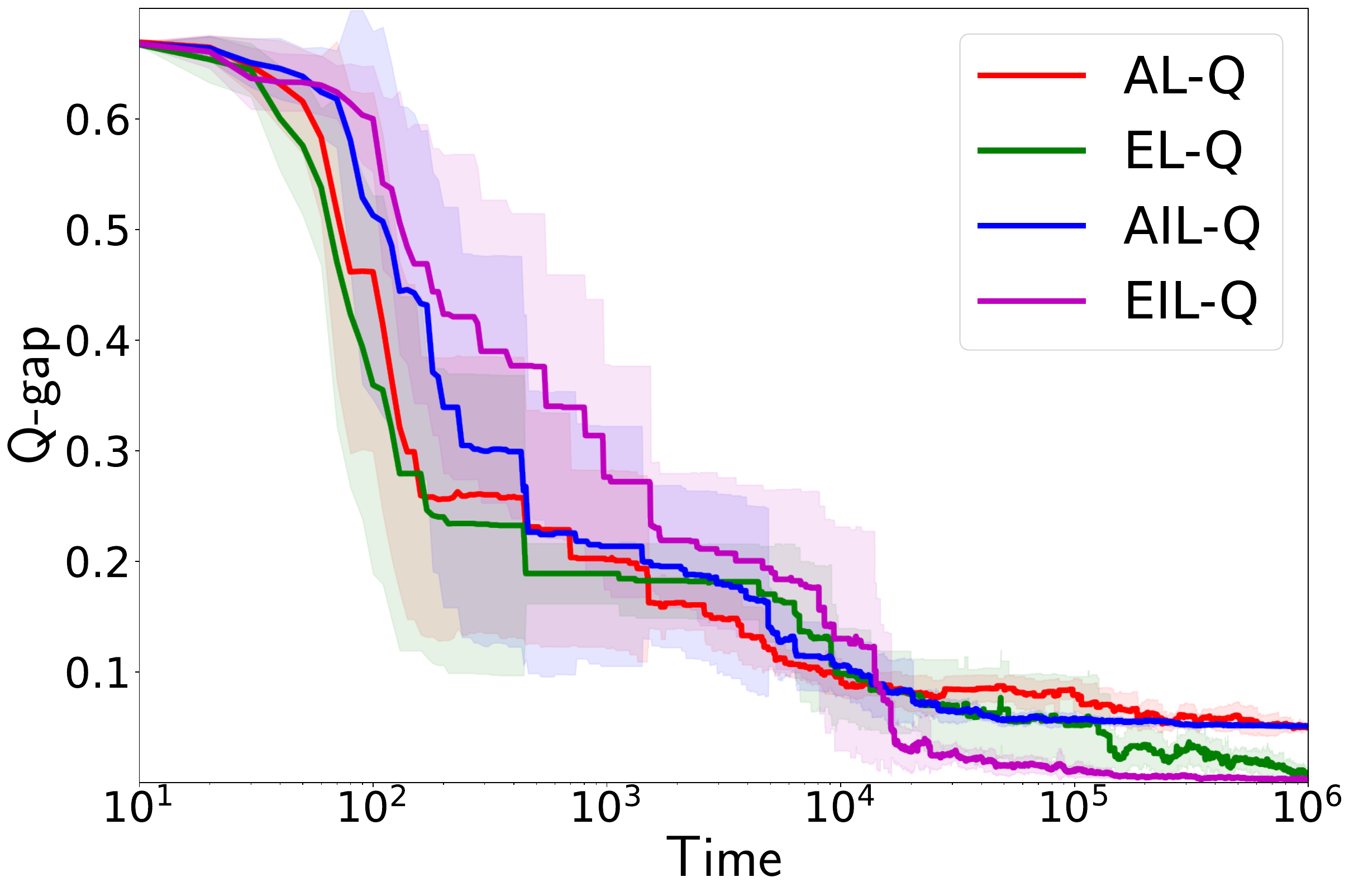}
            \caption{Q-gap vs Time}
    \color{black}
            \label{fig:QQstar}
        \end{subfigure}
    
        \begin{subfigure}[b]{0.525\textwidth}
    \color{black}
            \centering
    
            \includegraphics[width=\textwidth,page=2]{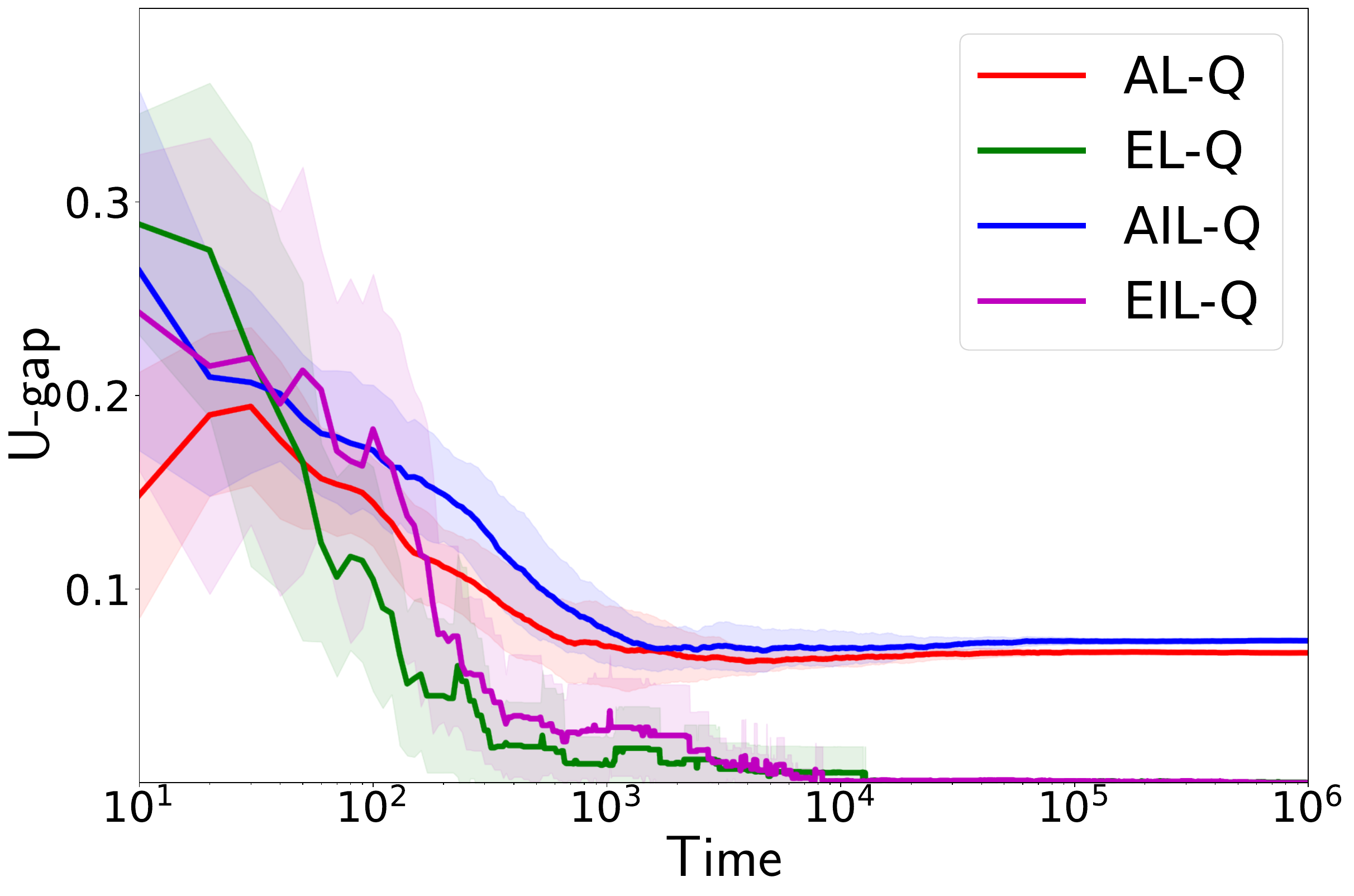}
            \caption{U-gap vs Time}
    \color{black}
            \label{fig:UUstar}
        \end{subfigure} 
    
        \begin{subfigure}[b]{0.525\textwidth}
    \color{black}
            \centering
    
            \includegraphics[width=\textwidth,page=3]{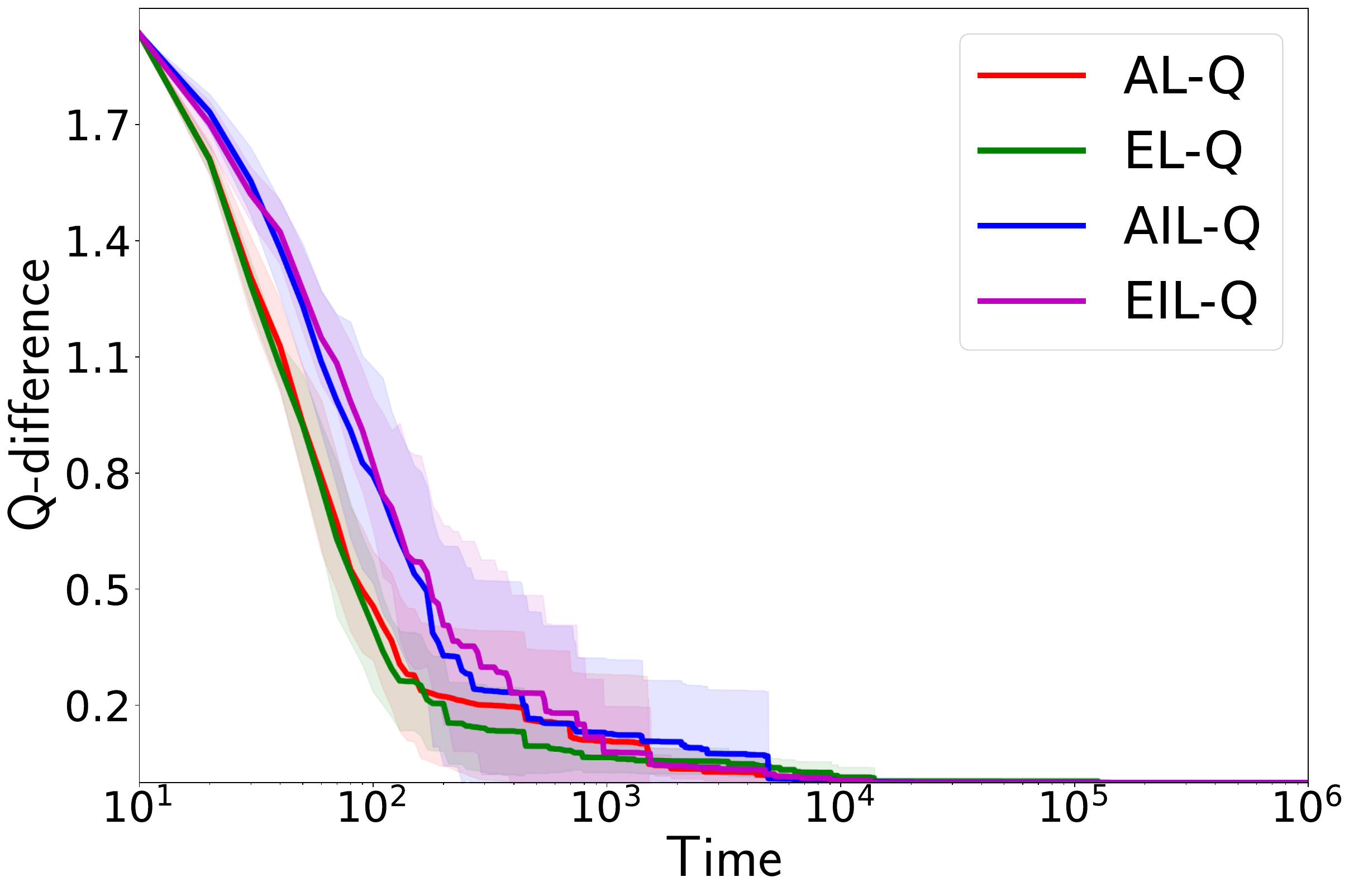}
            \caption{Q-difference vs Time}
    \color{black}
            \label{fig:QQ}
        \end{subfigure} 
        
        \caption{
          From top to bottom, the solid curves represent the AL-Q, EL-Q, AIL-Q, and EIL-Q dynamics, respectively. The shaded areas show the standard deviation across independent runs. In \Cref{fig:QQstar,fig:UUstar}, the gap for Q-functions and game values converge to the (near) optimal ones for all dynamics, corroborating \Cref{thm:main}. In \Cref{fig:QQ}, the impact of different initializations on the dynamics' long-run behavior diminishes in time, showing robustness to arbitrary initializations.}    
        \vspace{-3ex}
        \label{fig:simulation}
        \end{figure}
    
    \section{Illustrative Examples}\label{sec:examples} 
    In this section, we examine the long-run behavior of \Cref{alg} numerically. Consider a stochastic team with three states and three agents. Each agent can take two actions in each state (i.e., $|A^i|=2$ for each $i$ and $|A|=8$). The stage-payoffs are randomly selected so that Q-function estimates remain bounded within $[0,1]$. We set $\gamma=0.6$ and $\tau=0.05$ (and $\delta=0.2$ for the independent-revision schemes). For all agents, the initial Q-function values are assigned differently, e.g., $Q_0^i(s,a)=1+0.1\cdot i$ for all $(s,a)$, to examine the impact of different initializations. We run the experiments for $10^6$ time steps over $10$ independent trials. \Cref{fig:simulation} depicts both the mean error and standard deviation for these trials for the AL-Q, EL-Q, AIL-Q, and EIL-Q algorithms. We use a logarithmic scale for time to illustrate both the transient and steady-state behavior of the dynamics more explicitly.
    
    \Cref{fig:QQstar} shows the decay of \textit{Q-gap} (defined by $\|Q_t - Q_*\|_{\infty}$) to (near) zero, implying the convergence of the Q-function estimates $Q_t$ to the (near) optimal ones $Q_*$. Furthermore, \Cref{fig:UUstar} shows the decay of the \textit{U-gap} (defined by $\max_{\pi} \{U(\pi)\} - U(\pi_t))$ to (near) zero, implying the convergence to the (near) efficient equilibrium. Recall the averaging and exploration-free approaches for the estimated play, as described in \eqref{eq:piave}. In the AL-Q, and AIL-Q dynamics, using the averaging scheme, the Q-function estimates $Q_t$ and the estimated plays $\piave_t$ reach \textit{near} efficient equilibrium with an offset error induced by the exploration inherent to the logit dynamics to avoid convergence to inefficient equilibrium. On the other hand, in the EL-Q and EIL-Q dynamics, using the exploration-free scheme, the Q-function estimates $Q_t$ and the estimated plays $\pifree_t$ reach the \textit{exact} efficient equilibrium, mitigating the offset error.
    
    We further examine whether different initializations of Q-function estimates cause non-diminishing differences in their estimates in \Cref{fig:QQ}. We observe that the Q-difference (defined by $\sum_{i,j}\|Q_t^i - Q_t^j\|_{2}$) decays to zero for arbitrary initialization of $Q_0^i(s,a)=1+0.1\cdot i$ for all $(s,a)$.
    
    \section{Concluding Remarks}\label{sec:conclusion}
    This paper introduced a new family of logit-Q dynamics for efficient learning in \textit{unknown} stochastic teams by combining log-linear learning with Q-learning within the stage-game framework. We proposed four new dynamics: Averaged Logit-Q (AL-Q), Exploration-free Logit-Q (EL-Q), Averaged Independent Logit-Q (AIL-Q), and Exploration-free Independent Logit-Q (EIL-Q). These dynamics differ in terms of how agents revise their actions and model their joint play, offering flexibility for different learning scenarios. We showed that these dynamics converge to a (near) efficient equilibrium in stochastic teams, with quantifiable approximation errors. We also demonstrated the rationality of the dynamics against agents following pure stationary strategies and stochastic games with a single controller where the stage-payoffs specific to each state induces a potential game. We conducted various simulations to illustrate the convergence numerically. Future research could explore scaling these dynamics using function approximation or networked interconnections.

    \appendix
    \section{Proof of \Cref{lem:coupling}}\label{sec:coupling} By the coupling lemma \citep[Proposition 4.2, 4.7]{ref:Levin17}, we have:
    \[
    \|\mu_k^{w} - \what{\mu}_k\|_1 \;\leq\; 2 \,\Pr(w_{k} \neq \what{w}_{k}).
    \]
    We construct a coupling of the processes \(\{\what{w}_k\}\) and \(\{(w_k,z_k)\}\) to bound \(\Pr(w_{k} \neq \what{w}_{k})\). Let us describe the coupling in two cases:
    
    \textbf{Case (i):} \(\,w_k \neq \what{w}_k.\)
    The two processes evolve \emph{independently} under their respective transition kernels \(\mathbb{P}\) and \(\what{\mathbb{P}}\).
    
    \textbf{Case (ii):} \(\,w_k = \what{w}_k.\)
    They again evolve under \(\mathbb{P}\) and \(\what{\mathbb{P}}\), but now in a \emph{coupled} manner so that 
    \begin{align}
        \Pr\bigl(w_{k+1}\neq \what{w}_{k+1}\mid \what{w}_{k} = w_k , \{w_l,z_l \}_{l=0}^{k}\bigr)
        &= \TV{\mathbb{P}^{w}(\cdot \mid \{w_l,z_l \}_{l\leq k} )-\what{\mathbb{P}}(\cdot\mid w_k)}.
        \label{eq:coupled}
    \end{align}
    By Condition (ii), the total variation above is at most \(1-\lambda\).  Summing over any possible history \(\{z_l\}_{l\leq k}\) then yields, for all \(k\),
    \begin{align}
        \Pr\bigl(w_{k+1}\neq \what{w}_{k+1}\mid \what{w}_{k} = w_k , \{w_l\}_{l=0}^{k}\bigr) 
        \;\leq\; 1-\lambda.
        \label{eq:boundomegahistory}
    \end{align}
    
    \medskip
    \noindent\textbf{Bounding the mismatch probability.}  
    Condition (i) states that if \(w_k \neq \what{w}_k\), then with probability at least \(\varepsilon\,\what{\varepsilon}\) (over \(\kappa\) steps) both processes "meet" in the same state. Formally, there exist paths \(\{\what{h}^l\}_{l=0}^\kappa\), \(\{h^l\}_{l=0}^\kappa\) such that \(\what{h}^l \neq h^l\) for \(l<\kappa\) and \(\what{h}^\kappa = h^\kappa\). Hence,
    \begin{align}
    \Pr(w_{k+\kappa} = \what{w}_{k+\kappa} \;\mid\; w_k \neq \what{w}_k)
        \;\geq\; \varepsilon\,\what{\varepsilon},
        \label{eq:bound2}
    \end{align}
    where independence (Case (i)) ensures that probabilities $\varepsilon$ and, $\what{\varepsilon}$ can be multiplied for these two paths.
    
    We now partition time in blocks of length \(\kappa\).  From \eqref{eq:bound2} we have, whenever \(w_{(m-1)\kappa} \neq \what{w}_{(m-1)\kappa}\),
    \begin{equation}
    \Pr\bigl(w_{m\kappa} = \what{w}_{m\kappa} \;\mid\; w_{(m-1)\kappa} \neq \what{w}_{(m-1)\kappa}\bigr) 
    \;\ge\; \varepsilon\,\what{\varepsilon}, 
    \label{eq:condw1}
    \end{equation}
    while from \eqref{eq:boundomegahistory} and Case (ii), if \(w_{(m-1)\kappa} = \what{w}_{(m-1)\kappa}\) then 
    \begin{equation}
    \Pr\bigl(w_{m\kappa} = \what{w}_{m\kappa} \;\mid\; w_{(m-1)\kappa} = \what{w}_{(m-1)\kappa}\bigr) 
    \;\geq\; \lambda^{\kappa}. 
    \label{eq:condw3}
    \end{equation}
    Combining \eqref{eq:condw1}, \eqref{eq:condw3}, and using induction, one obtains for all \(m\ge0\):
    \begin{align}
        \Pr\bigl(w_{m\kappa} \neq \what{w}_{m\kappa}\bigr) 
        \;\le\; (1-\varepsilon\,\what{\varepsilon})^{m} \;+\; \frac{1-\lambda^{\kappa}}{\varepsilon\,\what{\varepsilon}}.
        \label{eq:ineq2}
    \end{align}

    Lastly, for any $k\in [m\kappa,(m+1)\kappa)$ and $m=0,1,\ldots$, the mismatch probability is bounded from above by
    \begin{align}
    \Pr(w_k \neq \what{w}_k) =&\; \Pr(w_{k} \neq \what{w}_{k}  \mid w_{m\kappa} \neq \what{w}_{m\kappa})\times \Pr(w_{m\kappa} \neq \what{w}_{m\kappa}) \nn\\
    &+ (1-\Pr(w_{k} = \what{w}_{k}  \mid w_{m\kappa} = \what{w}_{m\kappa}))\times\Pr(w_{m\kappa} = \what{w}_{m\kappa})\nn\\
    \leq&\; \Pr(w_{m\kappa} \neq \what{w}_{m\kappa}) +
    1 - \lambda^{\kappa}\label{eq:ineq1}
    \end{align}
    since $\Pr(w_{k} = \what{w}_{k}  \mid w_{m\kappa} = \what{w}_{m\kappa})\geq \lambda^{k-m\kappa} \geq \lambda^{\kappa}$ as in \eqref{eq:condw3}. Combining \eqref{eq:ineq2} and \eqref{eq:ineq1}, we obtain
    \begin{align}
    \Pr\left(w_k \neq \what{w}_{k}\right) \leq (1-\varepsilon\what{\varepsilon})^{m} +  (1-\lambda^{\kappa})\frac{1+\varepsilon\what{\varepsilon}}{\varepsilon\what{\varepsilon}}\quad \forall k\in [m\kappa,(m
    +1)\kappa).
    \end{align}
    for any $m=0,1,\ldots$. Since $1-\varepsilon\what{\varepsilon}\in(0,1)$, we have $(1-\varepsilon\what{\varepsilon})^{m} \leq (1-\varepsilon\what{\varepsilon})^{k/\kappa-1}$ for all $k\in [m\kappa,(m+1)\kappa)$. This completes the proof.
    

    \section{Proof of \Cref{lem:lambda}}\label{sec:lambda} We focus on the timesteps at which the state $s$ is visited.  Let $t_k$ be the time index of the $k$-th visit to $s$, and define $w_k := (s,a_{t_k})$ and $\what{w}_k := (s,\what{a}_{t_k})$.  Let $\mu_{k}^w,\,\what{\mu}_{k} \in \Delta(W)$ be their distributions at stage $t_k$, conditioned on $\mathcal{F}_{(n)}$.  
    
    The probability of transition from the action $\what{w}$ to $\what{w}'$ depends on $\what{w}$ and $Q_{(n)}$. Since $Q_{(n)}$ is $\mathcal{F}_{(n)}$-measurable, $\{\what{w}_{k}\}_k$ form a time homogeneous Markov chain conditioned on $\mathcal{F}_{(n)}$. Let $\what{\mathbb{P}}$ denote the transition matrix of this time homogeneous Markov chain. In contrast, $\{w_k\}_k$ is not time-homogeneous since $w_{k+1}$ depends on $w_k$ and on $Q_{t_{k+1}}(s,\cdot)$, which is not $\mathcal{F}_{(n)}$-measurable. Recall that \(z_k \in \mathbb{Z}\) denotes the trajectory of state-action pairs realized between the \(k\)-th and \((k+1)\)-th visits to state \(s\), where \(\mathbb{Z}\) is the countable set of all possible such trajectories. By the Q-update~\eqref{eq:Q}, each estimate \(Q_{t_{k+1}}(s,a)\) is \(\sigma(\{w_l,z_l\}_{l\le k})\)-measurable. Hence, the estimate \(Q_{(n)}\) (which is \(\mathcal{F}_{(n)}\)-measurable) and the history \(\{w_l,z_l\}_{l<k}\) from the \(\bigl(k_{(n)}-1\bigr)\)-th visitation to state \(s\) together determine the transition probability from \(w_{k-1} = (s,a_{t_{k-1}})\) to \(w_k = (s,a_{t_k})\). We denote these transition probabilities by \(\mathbb{P}\bigl(\cdot \mid \{w_l, z_l\}_{l\le k}\bigr)\). This implies that the discrete-time processes $\{\what{w}_{k}\}$ and $\{(w_k, z_k)\}$ are in the framework of \Cref{lem:coupling}. Correspondingly, in the following, we prove the claims that the conditions listed in \Cref{lem:coupling} hold so that we can invoke the lemma to address the distance between $\mu_{k}^{w}$ and $\what{\mu}_{k}$.
    
    \begin{claim}\label{claim:cond1}
    The processes $\{\what{w}_{k}\}$ and $\{ (w_k,z_k)\}$ satisfy Condition $(i)$ in \Cref{lem:coupling} for $\kappa = |A|$, $\varepsilon = (\epsilon\delta(1-\delta)^{n-1})^{2\kappa}$, and any $w'\in W$.
    \end{claim}
    \begin{proof}
    Due to the update rule of log-linear learning  $\what{w}=(s,\what{a}) \in W$ can transit to $\what{w}=(s,\what{a}') \in W$ only if $\what{a}$, and $\what{a}'$ are
    \begin{itemize}
    \item arbitrarily different at most for one agent's action in coordinated scheme,
    \item arbitrarily different for every agent's actions in independent scheme.
    \end{itemize}
    This yields that the Markov chain $\{\what{w}_{ k}\}_{k\geq k_{(n)} }$ is irreducible in both cases. The Markov chain also has self-loops for every extended state. 
    
    Since the Markov chain $\{\what{w}_{k}\}$ is irreducible and has self loops at every state, there exists a $\kappa$-length path $\what{w}=\what{h}_0,\ldots,\what{h}_{\kappa}=\what{w}'$ for any $(\what{w},\what{w}')\in W\times W$ in the fictional scenario, where $\kappa=|A|$. Even though $\{ (w_{k},z_k)\}$ do not form a homogeneous Markov chain, $\what{\mathbb{P}}$ and $\mathbb{P}$ have the same pattern. In other words, given any trajectory $\{w_l, z_{l}\}_{l < k}$, we have
    $\mathbb{P}^{w}(w'\mid , \{w_l, z_{l}\}_{l \leq k} ) > 0$ if and only if $\what{\mathbb{P}}(w'\mid w_k)>0$
    since agents updating their actions can take any action with some positive probability irrespective of the Q-function estimate due to the logit response \eqref{eq:logit} and the main and fictional scenarios differ only in terms of the Q-function estimates used.
    Therefore, there also exists a $\kappa$-length path $w = h^0,\ldots,h^{\kappa} = w'$ for any $(w,w')\in W\times W$ in the main scenario. 
    
    In the coordinated scheme, every feasible transition (in both main and fictional scenarios) has probability at least \(\tfrac{\epsilon}{n}\) because exactly one agent updates with probability \(\tfrac{1}{n}\) and then chooses an action with probability at least \(\epsilon\) (see \eqref{eq:minprob}).  These transitions are also valid in the independent scheme, where each such update occurs with probability at least \(\epsilon\,\delta\,(1-\delta)^{n-1}\).  Note that \(\delta\,(1-\delta)^{n-1} < \tfrac{1}{n}\).  Hence, in both settings, any \(\kappa\)-length path has probability at least \(\bigl(\epsilon\,\delta\,(1-\delta)^{n-1}\bigr)^{\kappa}\).  Therefore,
    \begin{subequations}\label{eq:path1}
        \begin{align}
        &\Pr\left(\bigwedge_{l=1}^{\kappa}\what{w}_{k+l}=\what{h}_{l}\mid \what{a}_{t_k} =\what{h}_0\right) \geq (\epsilon \delta(1-\delta)^{n-1})^{\kappa},\\
        &\Pr\left(\bigwedge_{l=1}^{\kappa}w_{ k+l}=h_l\mid (w_{k}, z_{k}) = (h_0, z) \right) \geq (\epsilon \delta(1-\delta)^{n-1})^{\kappa} \quad \forall z\in \mathbb{Z}.
        \end{align}
        \end{subequations}
    
    Given any \(w'\in W\) and any pair \(\bigl(\what{w}, w\bigr)\) with \(\what{w}\neq w\), there exist shortest paths of length at most \(\kappa\) from \(\what{w}\) to \(w'\) and from \(w\) to \(w'\), each changing at most one agent's action at a time.  If these paths share a common state, we align them from that point onward, inserting self-loops to ensure both paths have total length \(\kappa\). If one path is shorter (\(\ell < \what{\ell}\)), we let \(\{w_k\}\) follow its \(\ell\)-step path and then do self-loops at \(w'\) for \(\kappa-\ell\) steps, while \(\{\what{w}_k\}\) waits in self-loops at \(\what{w}\) for \(\kappa-\what{\ell}\) steps and then follows its \(\what{\ell}\)-step path (and vice versa).  If \(\ell = \what{\ell}\), each path still remains of length less than \(\kappa\), so we add self-loops at \(w'\) after traversing the path and add self-loops at \(\what{w}\) before traversing the path. In all cases, the two processes avoid occupying the same state simultaneously for the first \(\kappa\) steps, thanks to the artificially inserted delays, and each path has probability at least \(\bigl(\epsilon\,\delta\,(1-\delta)^{n-1}\bigr)^\kappa\) by \eqref{eq:path1}.  This establishes Condition~(i) of \Cref{lem:coupling} and completes the proof of \Cref{claim:cond1}.
    
    \end{proof}
    
    \begin{claim}\label{claim:cond2}
        The processes $\{\what{w}_{k} \}$ and $\{(w_{ k},z_k )\}$ meet Condition $(ii)$ in \Cref{lem:coupling} with $\lambda_{(n)}=1- CHK_{(n) }\times  \beta_{(n)} $, for some constant $C>0$, and sufficiently large $n$.
    \end{claim}
    \begin{proof}
    The transition kernels \(\mathbb{P}^w(\cdot\mid\{w_l,z_l\}_{l\le k})\) and \(\what{\mathbb{P}}(\cdot\mid \what{w}_k)\) both depend on agents' softmax responses to the Q-function estimates \(Q_{t_{k+1}}\) and \(Q_{(n)}\).  Since the softmax map~\eqref{eq:logit} is \(1/\tau\)-Lipschitz with respect to \(\|\cdot\|_2\) \citep[Proposition~4]{ref:Gao18}, there exists \(C>0\) such that
    \begin{align}\label{eq:cond}
        \TV{\what{\mathbb{P}}(\cdot\mid w_k) - \mathbb{P}^{w}(\cdot\mid \{w_l, z_{l}\}_{l \leq k})} \!\!&\leq C\|Q_{t_{k+1}} - Q_{(n)}\|_{\max} \leq  CH K_{(n)} \beta_{(n) },
        \end{align}
    for all $k\in [k_{(n)},k_{(n+1)})$, where $\beta_{(n)}:= \max_{a}\{\beta_{c_{(n)}(s,a)}\}$.  Hence we set $\lambda_{(n)}:=1 - C\,H\,K_{(n)}\,\beta_{(n)}$, which stays in $(0,1]$ for sufficiently large $n$ since $K_{(n)}\times\beta_{(n)}$ is decaying to zero as $n\rightarrow\infty$ by \Cref{lem:ane} whereas $C$, and $H$ are time-invariant. Hence, Condition $(ii)$ in \Cref{lem:coupling} also holds for sufficiently large $n$. This completes the proof of \Cref{claim:cond2}.
    \end{proof}
    
    By \Cref{claim:cond1,claim:cond2}, \Cref{lem:coupling} yields that
    \be\label{eq:lambdabound}
    \|\mu_{k}^{w} - \what{\mu}_{k}\|_1 \leq 2(1-\varepsilon)^{\frac{(k-k_{(n)})}{\kappa}-1} + 2\left(1-\lambda_{(n)}^{\kappa}\right)\frac{1+\varepsilon}{\varepsilon}
    \ee
    for all $k\geq k_{(n)}$. Recall that $\{\what{w}_k\}$ is irreducible and aperiodic, so it has a unique stationary distribution $\what{\mu}_{(n)}$. if we had  $\what{\mu}_{k_{(n)}-1} = \what{\mu}_{(n)}$, i.e., $\what{a}_{k_{(n)}}\sim \what{\mu}_{(n)}$, then we would have $\what{\mu}_{k} = \what{\mu}_{(n)}$, i.e., $\what{w}_{ k}\sim \what{\mu}_{(n)} $, for all $k\geq k_{(n)} $. Therefore, \eqref{eq:lambdabound} yields that
    \be\label{eq:lambdabound2}
    \|\mu_{ k}^{w} - \what{\mu}_{(n)}\|_1 \leq 2(1-\varepsilon)^{\frac{(k-k_{(n)} )}{\kappa}-1} + 2\left(1-\lambda_{(n)}^{\kappa}\right)\frac{1+\varepsilon}{\varepsilon}
    \ee
    for all $k\geq k_{(n)} $ which completes the proof of \Cref{lem:lambda}.

    \section{Proof of \Cref{lem:stepsizein}}\label{sec:stepsize}
    \color{black}
      The proof is by construction. First of all, by \eqref{eq:alphaLemma}, we have
      \begin{align}\label{eq:betaalpha}
      \alpha_{k_{(n+1)}-1} \leq \oalpha_{(n)} = \alpha_{k_{(n)}}\prod_{l=k_{(n)}+1}^{k_{(n+1)}-1}(1-\alpha_l) + \ldots + \alpha_{k_{(n+1)}-1} \leq K_{(n)} \alpha_{k_{(n)}}
      \end{align}
      since $\alpha_k$'s are monotonically non-increasing and take values in $[0,1]$. As $\alpha_k$ decays to zero, the bounds in \eqref{eq:betaalpha} would imply that $\oalpha_{(n)}$ also decays to zero if we had $\lim_{n\rightarrow\infty}K_{(n)} \alpha_{k_{(n)}} = 0 $ which is true if $K_{(n)}$ grows at a sub-linear rate, and $k_{(n)}\geq n$. Some algebra would also show that $\oalpha_{(n)} = 1-\prod_{k=k_{(n)}}^{k_{(n+1)}-1}(1-\alpha_k)$, and therefore, $\oalpha_{(n)}\in[0,1]$. Next we show that such a realization of $K_{(n)}$ is possible such that $\oalpha_{(n)}$ satisfies standard conditions.
      
      Define the following:
      \begin{flalign}
        \oA_N \coloneqq \sum_{k=1}^{N} \alpha_k, \quad \quad \uA_N \coloneqq \sum_{k=N}^{\infty} \alpha_k^2. 
      \end{flalign}
      By definition and the assumptions stated in the \Cref{lem:stepsizein}, $\lim_{N \rightarrow \infty }\oA_N = \infty$, and $\lim_{N \rightarrow \infty }\uA_N = 0$. Then, for any $m \in \mathbb{N}$,
      \begin{flalign}
        \onu(m) \coloneqq \min\{N: \oA_N \geq 2^m , N \in \mathbb{N}\}, \quad \unu(m) \coloneqq \min\{N: \uA_N \leq 2^{-m}, N \in \mathbb{N}\}
      \end{flalign}
      are well defined functions. $\onu(m)$ and $\unu(m)$ are monotonically non-decreasing by definition. Also, define the following:
      \begin{flalign}
        &\oK_{(n)} = \max(1, \max\{m: \oA_n \geq 2^m, m \in \mathbb{N}\}), \\ 
        &\uK_{(n)} = \max(1, \max\{m: \uA_n \leq 2^{-m}, m \in \mathbb{N}\}).
      \end{flalign}
      Note that, $\oK_{(n)} = m$ if and only if $\onu(m) \leq n < \onu(m+1)$, and $\uK_{(n)} = m$ if and only if $\unu(m) \leq n < \unu(m+1)$. Based on these definitons, we let $K_{(n)} = \min(\oK_{(n)}, \uK_{(n)})$. Notice that, by this definition $K_{(n)}$ grows at a sub-linear rate.
    
      Due to the epoch length having a minimum length of $1$, i.e. $K_{(n)} \geq 1$, and $k_{(1)}=1$, we have $k_{(n)} \geq n$. Morever, since $\alpha_k \in [0,1)$, we have $\onu(m) \geq m$ for all $m$.
    
      By the left inequality in \eqref{eq:betaalpha}, we have
      \begin{flalign}
        \sum_{n=1}^{\infty} \oalpha_{(n)} \geq \sum_{n=1}^{\infty} \alpha_{k_{(n+1)}-1} &\geq \sum_{n=1}^{\infty} \sum_{k=k_{(n+1)}-1}^{k_{(n+2)}-1}\frac{\alpha_k}{K_{(n+1)}} \\
        &\geq \sum_{k=k_{(2)}-1}^{\infty}\frac{\alpha_k}{K_{(k+1)}} \\
        &\geq \sum_{k=k_{(2)}-1}^{\infty}\frac{\alpha_k}{\oK_{(k+1)}} \\
        &\geq \sum_{m=k_{(2)}}^{\infty} \sum_{k = \onu(m)}^{\onu(m+1)-1} \frac{\alpha_k}{\oK_{(k+1)}} \\
        &\geq \sum_{m=k_{(2)}}^{\infty} \frac{1}{m+1} \left(\sum_{k = 1}^{\onu(m+1)}\alpha_k - \sum_{k = 1}^{\onu(m)-1} \alpha_k - 1\right) \\
        &\geq \sum_{m=k_{(2)}}^{\infty} \frac{1}{m+1} (2^{m+1}-2^m-1) \\
        &\geq \sum_{m=k_{(2)}}^{\infty} \frac{2^{m}-1}{m+1} = \infty.
      \end{flalign}
      
      By the right inequality in \eqref{eq:betaalpha}, we have
      \begin{flalign}
        \sum_{n=\unu(1)}^{\infty} \oalpha_{(n)}^2 \leq \sum_{n=\unu(1)}^{\infty} K_{(n)}^2\alpha_{k_{(n)}}^2 \leq \sum_{n=\unu(1)}^{\infty} \uK_{(n)}^2\alpha_{n}^2 &\leq \sum_{m=1}^{\infty} \sum_{n=\unu(m)}^{\unu(m+1)-1} \uK_{(n)}^2 \alpha_{n}^2 \\
        &\leq \sum_{m=1}^{\infty} (m)^2 \sum_{n=\unu(m)}^{\unu(m+1)-1} \alpha_{n}^2 \\
        &\leq \sum_{m=1}^{\infty} (m)^2 \sum_{n=\unu(m)}^{\infty} \alpha_{n}^2 \\
        &\leq \sum_{m=1}^{\infty} \frac{(m)^2}{2^{m}} \leq \infty.
      \end{flalign}
    For the claim that $K_{(n)}/k_{(n)}$ decaying to zero, we can simply use the definition of $K_{(n)}$. Since $\alpha_k< 1$ for all $k$, we can say that $\oA_{2^m}<2^m$. Then, $\oK_{(2^m)} < m$. Since $\{\oK_{(n)}\}_{n\geq1}$ is a monotonically non-decreasing sequence, $\oK_{(n)} \leq \lceil \log_2(n) \rceil$. Also, we know that $ k_{(n)} \geq n$. Then, the following is true
    \begin{flalign}
        0 \leq \lim_{n \rightarrow \infty} \frac{K_{(n)}}{k_{(n)}} \leq \lim_{n \rightarrow \infty} \frac{\oK_{(n)}}{n} \leq \lim_{n \rightarrow \infty} \frac{\lceil \log_2(n) \rceil}{n} = 0. 
    \end{flalign}
    This completes the proof.
    
    \section{Proof of \Cref{lem:stationary}}\label{sec:stationarydist}
    Let $a_t\in A$ and $\ua_t\in A$ denote the action profiles, resp., in the classical and independent log-linear dynamics. Both $\{a_t\}_{t\geq 0}$ and $\{\ua_t\}_{t\geq 0}$ form irreducible Markov chains with unique stationary distributions $\mu^{\mathrm{coor}}$ and $\mu^{\mathrm{ind}}$, respectively, and it is known that $\mu^{\mathrm{coor}} = \br(\Phi(\cdot))$.
    
    Let $P$ and $\uP$ denote the transition matrices in the Markov chains, resp., associated with the classical and independent log-linear learning dynamics. The ways agents update their actions in both dynamics yield that the transition matrices $P$ and $\uP$ satisfy
    $\uP = p_0 I+ p_1 P+ (1-p_0-p_1)E$,
    where $p_0 := (1-\delta)^n\in(0,1)$ and $p_1 := n\delta(1-\delta)^{n-1}\in (0,1)$ correspond, resp., to the probability that none and one of the agents update their actions, and $E$ is the transition matrix corresponding to the case at least two (randomly picked) agents updates their actions. The transition matrix corresponding to the case that at least one (randomly picked) agent updates their actions is given by 
    \be \label{eq:mat_id}
    \wbar{P} = \frac{p_1}{1-p_0}P+\frac{1-p_0-p_1}{1-p_0}E.
    \ee
    Since $\wbar{P}$ and $\uP$ share the same unique stationary distribution, we focus on ${\wbar{a}_t}$ evolving by $\wbar{P}$ rather than $\uP$, and couple its evolution with ${a_t}$ based on \Cref{lem:coupling}
    
    To apply \Cref{lem:coupling}, we verify two conditions. First, note that for any two states $a,\wbar{a} \in A$, both dynamics allow transitions in which only one agent updates at a time. Thus, there exist $n$-length paths from $a$ and from $\wbar{a}$ to a common state (say, $a'$) with consecutive profiles differing in at most one coordinate while both process differ until they reach $a'$ at step $n$. Since each one-agent update occurs with probability uniformly bounded below by a constant $\epsilon>0$, the probability of following any such $n$-step path is at least $\left(\frac{\epsilon}{N}\right)^n$ for the process $\{a_t\}$ and $\left(\frac{\epsilon\delta(1-\delta)^{n-1}}{1-(1-\delta)^n}\right)^{n}$ for the process $\{\wbar{a}_t\}$, establishing the required path connectivity.
    
    On the other hand, \eqref{eq:mat_id} yields that
    \begin{align}
    \TV{P(\cdot\mid h) - \wbar{P}(\cdot\mid  h)} &=\frac{1-p_0-p_1}{1-p_0} \cdot \TV{  P(\cdot\mid  h) - E(\cdot\mid  h) }\label{eq:TV_dev}
    \end{align}
    for any $h \in A$. Note that $\TV{  P(\cdot\mid  h) - E(\cdot\mid  h) }\leq 1$. 
    Therefore, Condition $(ii)$ in \Cref{lem:coupling} holds for $\lambda = 1- (1-p_0-p_1)/(1-p_0) = p_1/(1-p_0)\leq 1$. Since Conditions $(i)$ and $(ii)$ hold, we can invoke \Cref{lem:coupling} as let $t\rightarrow\infty$ to conclude \eqref{eq:resultstationary}. 
    
\begin{spacing}{1}
\bibliographystyle{plainnat}

\bibliography{mybib}
\end{spacing}
    
\end{document}